\documentclass[lettersize,journal]{IEEEtran}

\ifCLASSOPTIONcompsoc
  \usepackage[nocompress]{cite}
\else
  \usepackage{cite}
\fi

\usepackage[caption=false,font=footnotesize]{subfig}
\usepackage[export]{adjustbox}
\usepackage{tabularx}
\usepackage{colortbl}
\usepackage{multirow}
\usepackage{stmaryrd}
\usepackage{arydshln}
\usepackage{amsmath,amssymb}
\usepackage{mathtools}
\usepackage{formal-grammar}
\usepackage{tikz}
\usetikzlibrary{arrows,fit,positioning,shapes,patterns}
\usepackage{pgfplots}
\usepackage{forest}
\usepackage{soul}
\usepackage{listings}
\usepackage{array}
\usepackage{booktabs}
\pgfplotsset{width=8cm,compat=1.9}
\definecolor{scgreen}{HTML}{7fc97f}
\definecolor{scorange}{HTML}{fdc086}
\definecolor{scpurple}{HTML}{beaed4}
\definecolor{scred}{HTML}{e41a1c}
\definecolor{scdgreen}{HTML}{4daf4a}

\newcommand{\egg}{\texttt{egg}}

\newcommand{\mycc}{\cellcolor{lightgray}}
\usepackage{soul}

\newcommand*\change[1]{#1}
\newcommand*\avgareavsls{17.6\%}
\newcommand*\avgareavsrov{55.8\%}
\newcommand*\bst[1]{{\bf #1}}
\begin{document}
\title{ROVER: RTL Optimization via Verified\\ E-Graph Rewriting}

\author{Samuel~Coward,
        Theo~Drane,
        and~George~A.~Constantinides,~\IEEEmembership{Senior Member,~IEEE,}% <-this % stops a space
\IEEEcompsocitemizethanks{\IEEEcompsocthanksitem S.Coward and T.Drane are with the Intel Numerical Hardware Group.\protect\\
E-mail: samuel.coward@intel.com
\IEEEcompsocthanksitem G.A.Constantinides and S.Coward are with Imperial College London, Department of Electronic and Electrical Engineering, London, UK}% <-this % stops an unwanted space
\thanks{Manuscript received November 17, 2023}}

\maketitle

% \IEEEtitleabstractindextext{%
\begin{abstract}

Manual RTL design and optimization remains prevalent across the semiconductor industry because commercial logic and high-level synthesis tools are unable to match human designs. Our experience in industrial datapath design demonstrates that manual optimization can typically be decomposed into a sequence of local equivalence preserving transformations. 
By formulating datapath optimization as a graph rewriting problem we automate design space exploration in a tool we call ROVER. 

We develop a set of mixed precision RTL rewrite rules inspired by designers at Intel and an accompanying automated validation framework. A particular challenge in datapath design is to determine a productive order in which to apply transformations as this can be design dependent. ROVER resolves this problem by building upon the e-graph data structure, which compactly represents a design space of equivalent implementations. By applying rewrites to this data structure, ROVER generates a set of efficient and functionally equivalent design options. From the ROVER generated e-graph we select an efficient implementation. To accurately model the circuit area we develop a theoretical cost metric and then an integer linear programming model to extract the optimal implementation. To build trust in the generated design ROVER also produces a back-end verification certificate that can be checked using industrial tools. 

We apply ROVER to both Intel-provided and open-source benchmarks, and see up to a 63\% reduction in circuit area. ROVER is also able to generate a customized library of distinct implementations from a given parameterizable RTL design, improving circuit area across the range of possible instantiations. 

%\gc{I feel like the abstract doesn't necessarily capture the novelty, more the impact. For example, I guess we should say in here something about there being a traditional problem with knowing which order to apply rewrites in, and the e-graph setting solving this problem. We also say nothing about the techniques we've applied that are not just direct applications of egg, e.g. conditions for parametric RTL, nodes modelling behaviour of commercial RTL tools, ILP extraction, etc. I'm mainly worried that someone reading the abstract alone will think the story is: ``We applied this tool from other people to this problem. It worked well.''}
\end{abstract}

\begin{IEEEkeywords}
hardware optimization, design automation, datapath design, computer arithmetic
\end{IEEEkeywords}%}

% \IEEEdisplaynontitleabstractindextext

% \fi
% \IEEEpeerreviewmaketitle

\section{Introduction}\label{sec:introduction}

\IEEEPARstart{I}{n} recent years many new domain specific languages and tools have allowed hardware engineers to write designs at different levels of abstraction~\cite{Canis2013LegUp:Systems,Cadence2023StratusHLS}. Even with these developments, Register Transfer Level (RTL) design using hardware description languages such as Verilog still dominates industry and much of academia. Despite reaching maturity, logic and high-level synthesis tools are limited in their design space exploration and are unable to match skilled engineers. The hardware design space is large and mostly unexplored due to strict correctness requirements and slow debug time frames. The numerous optimization objectives and constraints present a challenge for both humans and automated systems. Automatic datapath synthesis research has focused on heuristic search and statistical methods~\cite{dataflow2008verma,Xydis2012Compiler-in-the-loopTrade-offs,HarishRam2012ADatapaths,Krishnan2006ASynthesis} or deployed machine learning~\cite{Roy2021PrefixRL:Learning}. Automatic datapath synthesis can expand design space exploration resulting in better quality circuit designs. It may also improve productivity reducing the engineering effort required to produce an optimized implementation. 

\begin{figure}
    \centering
    \begin{tikzpicture}
\begin{axis}[
    xlabel={Proportion of Rewrites Applied},
    ylabel={Circuit Area Metric Change (\%)},
    xmin=0, xmax=1.05,
    % ymode=log,
    % ymin=0, ymax=120,
    % xtick={0,20,40,60,80,100},
    % ytick={0,20,40,60,80,100,120},
    legend pos=north west,
    % ymajorgrids=true,
    % grid style=dashed,
]

% \addplot[
%     color=blue,
%     % mark=square,
%     ]
%     coordinates {
% (0,1)
% (0.016393442,1.0637165)
% (0.032786883,1.0958993)
% (0.04918033,1.2952245)
% (0.06557377,1.2630417)
% (0.08196721,1.2810796)
% (0.09836066,1.2810796)
% (0.114754096,1.2810796)
% (0.13114753,1.2726446)
% (0.14754099,1.2767973)
% (0.16393442,1.2767973)
% (0.18032786,1.4719698)
% (0.19672132,1.689333)
% (0.21311475,1.689333)
% (0.22950819,1.689333)
% (0.24590164,1.680898)
% (0.26229507,1.7412406)
% (0.27868852,1.7734233)
% (0.29508197,1.9727485)
% (0.3114754,1.9405658)
% (0.32786885,1.9586037)
% (0.3442623,1.9586037)
% (0.36065573,1.9586037)
% (0.37704918,1.9501687)
% (0.39344263,1.7453932)
% (0.40983605,1.7453932)
% (0.4262295,1.9405658)
% (0.44262296,2.157929)
% (0.45901638,2.157929)
% (0.47540984,2.157929)
% (0.4918033,2.149494)
% (0.5081967,1.7412406)
% (0.52459013,1.7734233)
% (0.5409836,1.9727485)
% (0.55737704,1.9405658)
% (0.57377046,1.9586037)
% (0.59016395,1.9586037)
% (0.60655737,1.9586037)
% (0.6229508,1.9501687)
% (0.6393443,1.7453932)
% (0.6557377,1.7453932)
% (0.6721311,1.9405658)
% (0.6885246,2.157929)
% (0.704918,2.157929)
% (0.72131145,2.157929)
% (0.73770493,2.149494)
% (0.75409836,1.6542953)
% (0.7704918,1.6542953)
% (0.78688526,1.6542953)
% (0.8032787,1.6221126)
% (0.8196721,1.6305476)
% (0.8360656,1.6305476)
% (0.852459,1.6305476)
% (0.86885244,1.6221126)
% (0.8852459,1.4173372)
% (0.90163934,1.4173372)
% (0.91803277,1.3810018)
% (0.93442625,1.3894368)
% (0.9508197,1.3894368)
% (0.9672131,1.3894368)
% (0.9836066,1.3810018)
% (1,0.9090319)
%     };

\addplot[
    color=gray,
    dashed
    ]
    coordinates {
    (-0.1,0)
    (1.1,0)};

% based on dg_weight rewriting
\addplot[
    color=red,
    ]
    coordinates {
(0,           0)
(0.0061728396,121.12725-100)
(0.012345679, 121.78423-100)
(0.018518519, 122.44122-100)
(0.024691358, 122.44122-100)
(0.030864198, 122.44122-100)
(0.037037037, 121.78423-100)
(0.043209877, 122.44122-100)
(0.049382716, 123.09821-100)
(0.055555556, 122.44122-100)
(0.061728396, 123.09821-100)
(0.06790123,  122.44122-100)
(0.074074075, 121.78423-100)
(0.08024691,  121.78423-100)
(0.086419754, 121.78423-100)
(0.09259259,  121.78423-100)
(0.09876543,  121.78423-100)
(0.10493827,  122.18188-100)
(0.11111111,  121.78423-100)
(0.11728395,  122.44122-100)
(0.12345679,  123.09821-100)
(0.12962963,  123.09821-100)
(0.13580246,  123.09821-100)
(0.14197531,  122.44122-100)
(0.14814815,  123.09821-100)
(0.15432099,  123.75518-100)
(0.16049382,  123.09821-100)
(0.16666667,  123.75518-100)
(0.17283951,  123.09821-100)
(0.17901234,  122.44122-100)
(0.18518518,  123.09821-100)
(0.19135803,  123.09821-100)
(0.19753087,  123.09821-100)
(0.2037037,   122.44122-100)
(0.20987654,  124.44675-100)
(0.21604939,  124.44675-100)
(0.22222222,  124.44675-100)
(0.22839506,  122.44122-100)
(0.2345679,   122.73513-100)
(0.24074075,  122.44122-100)
(0.24691358,  143.56847-100)
(0.25308642,  144.22545-100)
(0.25925925,  144.88243-100)
(0.2654321,   144.88243-100)
(0.27160493,  144.88243-100)
(0.2777778,   144.22545-100)
(0.28395063,  144.88243-100)
(0.29012346,  145.53941-100)
(0.2962963,   144.88243-100)
(0.30246913,  145.53941-100)
(0.30864197,  144.88243-100)
(0.3148148,   144.22545-100)
(0.32098764,  144.22545-100)
(0.3271605,   144.22545-100)
(0.33333334,  144.22545-100)
(0.33950618,  144.22545-100)
(0.34567901,  144.6231-100)
(0.35185185,  144.22545-100)
(0.3580247,   144.22545-100)
(0.36419752,  144.88243-100)
(0.37037036,  144.88243-100)
(0.37654322,  144.88243-100)
(0.38271606,  144.22545-100)
(0.3888889,   144.88243-100)
(0.39506173,  145.53941-100)
(0.40123457,  144.88243-100)
(0.4074074,   145.53941-100)
(0.41358024,  144.88243-100)
(0.41975307,  144.22545-100)
(0.42592594,  144.88243-100)
(0.43209878,  144.88243-100)
(0.4382716,   144.88243-100)
(0.44444445,  144.22545-100)
(0.45061728,  145.7296-100)
(0.45679012,  145.7296-100)
(0.46296296,  147.73513-100)
(0.4691358,   147.73513-100)
(0.47530866,  145.7296-100)
(0.4814815,   124.96542-100)
(0.48765433,  125.6224-100)
(0.49382716,  126.27939-100)
(0.5,         126.27939-100)
(0.50617284,  126.27939-100)
(0.5123457,   125.6224-100)
(0.5185185,   126.27939-100)
(0.52469134,  126.93638-100)
(0.5308642,   126.27939-100)
(0.537037,    126.93638-100)
(0.54320985,  126.27939-100)
(0.5493827,   125.6224-100)
(0.5555556,   125.6224-100)
(0.5617284,   125.6224-100)
(0.56790125,  125.6224-100)
(0.5740741,   125.6224-100)
(0.5802469,   126.02005-100)
(0.58641976,  125.6224-100)
(0.5925926,   125.6224-100)
(0.59876543,  126.27939-100)
(0.60493827,  126.27939-100)
(0.6111111,   126.27939-100)
(0.61728394,  125.6224-100)
(0.6234568,   126.27939-100)
(0.6296296,   126.93638-100)
(0.63580245,  126.27939-100)
(0.6419753,   126.93638-100)
(0.6481481,   126.27939-100)
(0.654321,    125.6224-100)
(0.66049385,  126.27939-100)
(0.6666667,   126.27939-100)
(0.6728395,   126.27939-100)
(0.67901236,  125.6224-100)
(0.6851852,   126.88451-100)
(0.69135803,  126.88451-100)
(0.69753087,  126.88451-100)
(0.7037037,   124.87898-100)
(0.70987654,  103.83817-100)
(0.7160494,   103.54426-100)
(0.7222222,   103.54426-100)
(0.72839504,  103.54426-100)
(0.7345679,   103.83817-100)
(0.7407407,   103.83817-100)
(0.74691355,  103.83817-100)
(0.75308645,  103.83817-100)
(0.7592593,   103.54426-100)
(0.7654321,   082.78008-100)
(0.77160496,  082.78008-100)
(0.7777778,   083.43707-100)
(0.7839506,   083.43707-100)
(0.79012346,  083.43707-100)
(0.7962963,   082.78008-100)
(0.80246913,  082.78008-100)
(0.80864197,  083.2296-100)
(0.8148148,   082.572615-100)
(0.82098764,  083.2296-100)
(0.8271605,   082.572615-100)
(0.8333333,   081.91563-100)
(0.83950615,  081.91563-100)
(0.845679,    081.91563-100)
(0.8518519,   081.91563-100)
(0.8580247,   081.91563-100)
(0.86419755,  082.313275-100)
(0.8703704,   081.91563-100)
(0.8765432,   081.25864-100)
(0.88271606,  081.91563-100)
(0.8888889,   081.91563-100)
(0.89506173,  081.91563-100)
(0.90123457,  081.25864-100)
(0.9074074,   081.25864-100)
(0.91358024,  081.708163-100)
(0.9197531,   081.051177-100)
(0.9259259,   081.708163-100)
(0.93209875,  081.051177-100)
(0.9382716,   080.39419-100)
(0.9444444,   081.051177-100)
(0.9506173,   081.051177-100)
(0.95679015,  081.051177-100)
(0.962963,    080.39419-100)
(0.9691358,   081.656295-100)
(0.97530866,  081.656295-100)
(0.9814815,   079.65076-100)
(0.9876543,   079.65076-100)
(0.99382716,  077.64523-100)
(1,           053.78631-100)
};

\end{axis}
\end{tikzpicture}
    \caption{\change{Progression of design cost throughout RTL rewriting for the Weight Calculation benchmark (described in Section~\ref{sect:results}).} We plot the percentage change in the circuit area metric compared to the original design at every point in the rewrite chain. The area metric may converge non-monotonically.}
    \label{fig:rewrite_vs_cost}
\end{figure}
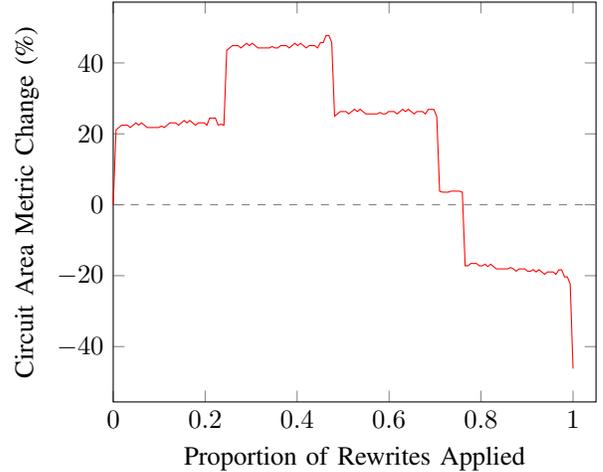

Inspired by traditional compiler optimization techniques and previous work on RTL optimization~\cite{dataflow2008verma,Cooper2011EngineeringEdition}, we observe that manual datapath optimization at RTL can be described in terms of local equivalence-preserving transformations. Skilled engineers learn such transformations through experience and discover patterns or sequences of valuable transformations. Often these optimizations can be generalized, facilitating their application more widely. Automating transformation-driven hardware optimization is complex since it is often necessary to apply several ``bad'' transformations before an ultimately beneficial transformation can be applied. \change{Figure~\ref{fig:rewrite_vs_cost} illustrates an example where it is necessary to initially apply transformations that increase circuit area cost via operator duplication or replacement, but eventually lead to subsequent area saving transformations such as arithmetic simplification or clustering, providing a net area reduction.} This is a challenge faced by traditional rewriting techniques~\cite{Verma2008VariableDesign}. 

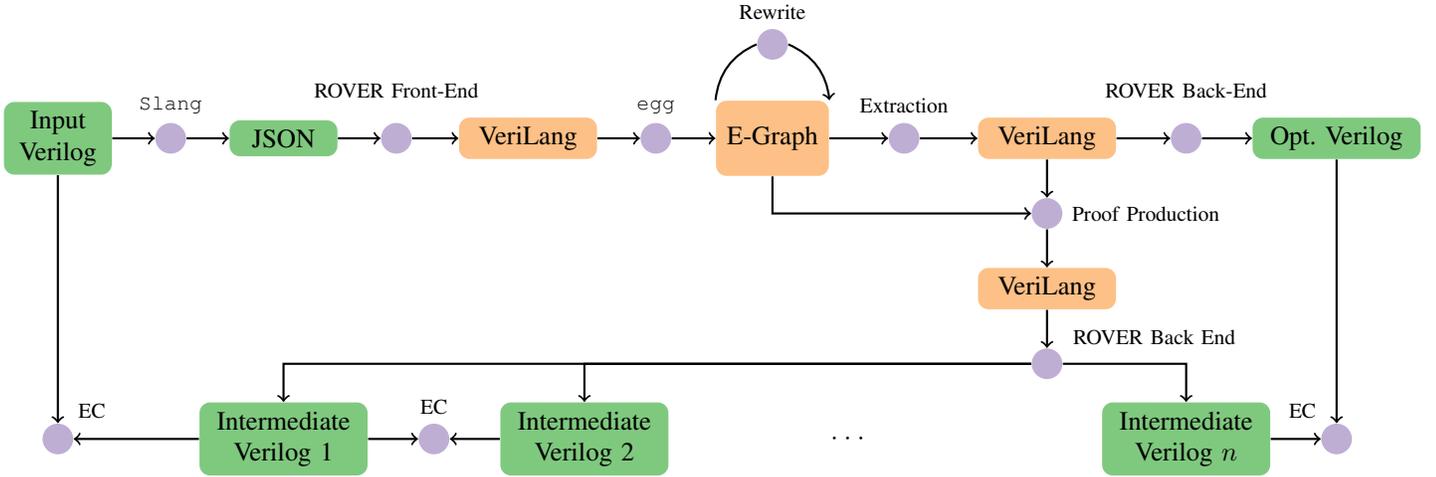
\begin{figure*}
    \centering
    \begin{tikzpicture}

% Boxes
\node [shape=rectangle, fill=scgreen, text width = 1.2cm, text centered,rounded corners] at (-8,0) (input) {Input Verilog};
\node [shape=circle, fill=scpurple, text width = 0.1cm, text centered,label={\footnotesize{\texttt{Slang}}}] at (-6.5,0) (slang) {};
\node [shape=rectangle, fill=scgreen, text width = 1.2cm, text centered,rounded corners] at (-5,0) (json) {JSON};
\node [shape=circle, fill=scpurple, text width = 0.1cm, text centered,label={[label distance=0.2cm]\footnotesize{ROVER Front-End}}] at (-3.5,0) (front_end) {};
\node [shape=rectangle, fill=scorange, text width = 1.6cm, text centered,rounded corners] at (-1.75,0) (verilang) {VeriLang};

\node [shape=circle, fill=scpurple, text width = 0.1cm, text centered,label={\footnotesize{\egg}}] at (-0.05,0) (egg) {};
\node [shape=rectangle, fill=scorange, minimum width=1.5cm, minimum height=1cm,rounded corners] at (1.5,0) (egraph) {E-Graph};

\node [shape=circle, fill=scpurple, text width = 0.1cm, text centered,label={\footnotesize{Rewrite}}] at (1.5,1.25) (rewrites) {};
\node [shape=circle, fill=scpurple, text width = 0.1cm, text centered,label={\footnotesize{Extraction}}] at (3.25,0) (extract) {};
\node [shape=rectangle, fill=scorange, text width = 1.6cm, text centered, rounded corners] at (5.15,0) (opt_verilang) {VeriLang};
\node [shape=circle, fill=scpurple, text width = 0.1cm, text centered,label={[label distance=0.2cm]\footnotesize{ROVER Back-End}}] at (7,0) (back_end) {};
\node [shape=rectangle, fill=scgreen, text width = 2cm, text centered, rounded corners] at (9,0) (output) {Opt. Verilog};

\node [shape=circle, fill=scpurple, text width = 0.1cm, text centered,label=0:{\footnotesize{Proof Production}}] at (5.15,-1) (proof_prod) {};

\node [shape=rectangle, fill=scorange, text width = 1.6cm, text centered, rounded corners] at (5.15,-2) (proof) {VeriLang};

\node [shape=circle, fill=scpurple, text width = 0.1cm, text centered,label={[label distance=0.05cm]30:\footnotesize{ROVER Back End}}] at (5.15,-3) (back_end_proof) {};
\node [shape=rectangle, fill=scgreen, text width = 2cm, text centered,rounded corners] at (-5,-4) (int1) {Intermediate Verilog 1};
\node [shape=rectangle, fill=scgreen, text width = 2cm, text centered,rounded corners] at (-1,-4) (int2) {Intermediate Verilog 2};
\node [shape=ellipse] at (2.5,-4) (dots) {$\cdot \cdot \cdot$};
\node [shape=rectangle, fill=scgreen, text width = 2cm, text centered,rounded corners] at (7,-4) (intn) {Intermediate Verilog $n$};

\node [shape=circle, fill=scpurple, text width = 0.1cm, text centered,label=45:{\footnotesize{EC}}] at (-8,-4) (fv_0) {};

\node [shape=circle, fill=scpurple, text width = 0.1cm, text centered,label={\footnotesize{EC}}] at (-3,-4) (fv_1) {};

\node [shape=circle, fill=scpurple, text width = 0.1cm, text centered,label=135:{\footnotesize{EC}}] at (9,-4) (fv_2) {};

\draw [->,thick] (input) edge (slang);
\draw [->, thick] (slang) edge (json);
\draw [->, thick] (json) edge (front_end);
\draw [->, thick] (front_end) edge (verilang);
\draw [->, thick] (verilang) edge (egg);
\draw [->, thick] (egg) edge (egraph);
\path [->, thick] (rewrites.east) edge[bend left] (egraph.north east);
\draw [-, thick] (egraph.north west) edge[bend left] (rewrites.west);

\draw [->, thick] (egraph) edge (extract);
\draw [->, thick] (egraph) |- (proof_prod);
\draw [->, thick] (extract) edge (opt_verilang);
\draw [->, thick] (opt_verilang) edge (back_end);
\draw [->, thick] (back_end) edge (output);

\draw [->,  thick] (opt_verilang) edge (proof_prod);
\draw [->,  thick] (proof_prod) edge (proof);
\draw [->,  thick] (proof) edge (back_end_proof);
\draw [->,  thick] (back_end_proof) -| (int1);
\draw [->,  thick] (back_end_proof) -| (int2);
\draw [->,  thick] (back_end_proof) -| (intn);
\draw[->,  thick] (input) edge (fv_0);
\draw[->,  thick] (int1) edge (fv_0);
\draw[->,  thick] (int1) edge (fv_1);
\draw[->,  thick] (int2) edge (fv_1);

\draw[->,  thick] (intn) edge (fv_2);
\draw[->,  thick] (output) edge (fv_2);

\end{tikzpicture}
    \caption{Flow diagram describing the operation of ROVER. The intermediate RTL designs are formally verified to be functionally equivalent using a commercial equivalence checker (EC) forming a chain of reasoning. The orange boxes denote the novel contributions.}
    \label{fig:flow_diagram}
\end{figure*}

In order to meet the automation objective, we leverage recent advances in e(quivalence)-graph rewriting and equality saturation, bringing them to the RTL optimization problem. By representing combinational RTL as a dataflow graph we can exploit properties of e-graphs that make them a promising technology for hardware design. Firstly, in e-graph rewriting the order in which transformations are applied is unimportant, allowing the e-graph to capture early transformations that initially degrade the design but potentially enable later beneficial optimizations. Secondly, e-graphs are designed to explore equivalent implementations, and RTL optimization typically maintains functionality producing bit-identical implementations. Lastly, the e-graph maintains the complete history of all designs it has explored, which allows us to decompose formal verification into a sequence of equivalence checks. 
% An introduction to e-graphs and equality saturation is given in Section~\ref{subsec:egraphs}.

In this paper, we address the following problem. Given an RTL implementation $R$, we seek a functionally equivalent implementation $R'$ that minimizes some cost, typically area or delay. Two implementations $R$ and $R'$ are equivalent, $R\cong R'$, iff for all inputs they produce identical outputs. The resulting optimized implementation $R'$ is passed to an industrial logic synthesis tool, producing a netlist from which we can extract relevant circuit quality metrics. ROVER's optimization objective is to generate RTL that the logic synthesis tool can synthesize into the most efficient circuit representation. This means that ROVER must capture and model the downstream logic synthesis capabilities. Figure~\ref{fig:flow_diagram} provides an overview of the tool flow. 

The primary contributions of this work are:
\begin{itemize}
    \item application of e-graph rewriting to RTL datapath optimization,
    \item a multi-bitwidth and multi-signage rewrite set that enables datapath design space exploration capturing the connection between optimal architecture selection and bitwidth,
    \item an automated method to generate necessary and sufficient conditions for RTL rewrites using an equivalence checker,
    \item a robust method to verify the correctness of the generated RTL based on problem decomposition.
    % \item an efficient linear programming based extraction method.
\end{itemize}

An initial application of e-graphs to general datapath optimization was presented at the 29th Symposium on Computer Arithmetic~\cite{Coward2022AutomaticE-Graphs} as a preliminary version of this work. Here we extend the conference paper by supporting signed arithmetic, providing semantics for ROVER's intermediate language, and introducing novel methods for rewrite condition generation and formal verification not present in the conference version.

In the next section we provide the necessary background on datapath optimization and e-graphs. In Section~\ref{sect:language} we describe the intermediate language and supported subset of Verilog. Then we describe the rewrites that encode the optimizations and allow the e-graph to grow in Section \ref{sect:rewrites}. In Section~\ref{sect:extraction} we describe how the optimal design is extracted from the generated e-graph. The verification methodology is described in Section~\ref{sect:verification}. In the final two sections we present results.

%%%%%%%%%%%%%%%%%%%%%%%%%%%%%%%%%%%%%%%%%%%%%%%%%%%%%%%%%%%%%%%%%
% BACKGROUND
%%%%%%%%%%%%%%%%%%%%%%%%%%%%%%%%%%%%%%%%%%%%%%%%%%%%%%%%%%%%%%%%%
\section{Background} \label{sect:background}
\subsection{Datapath Synthesis} \label{subsect:datapath_opto}
Datapath synthesis is the process of generating gate level netlists from higher-level arithmetic circuit designs expressed in RTL. Zimmermann decomposes this process into three steps: RTL  extraction of arithmetic operations, followed by high-level arithmetic optimizations, and finally netlist generation~\cite{Zimmermann2009DatapathDesign}. Such datapath optimization engines form a core component of all logic and high-level synthesis tools and are essential for generating state-of-the-art circuit designs.

Logic synthesis tools implement a range of hardware-specific optimizations, detecting opportunities to merge particular operator sequences and exploit redundant number representations~\cite{Zimmermann2009DatapathDesign}. Synopsys Design Compiler provides datapath coding guidelines, which describe how designers can best exploit the synthesis tool's capabilities~\cite{Synopsys2019CodingSynthesis}. Of particular relevance is the front-end logic synthesis pass that performs datapath extraction, which clusters operators into datapath blocks~\cite{Zimmermann2009DatapathDesign}. Extracting larger clusters enables more effective downstream optimization. \change{Datapath clustering can be prevented by datapath leakage in a design, where a designer, possibly intentionally, truncates an arithmetic operation.} A key objective of ROVER is to enable logic synthesis datapath extraction to form larger datapath blocks, which in turn results in more efficient circuit implementations. 

% In the HLS \gc{I don't think you've used this acronym before? Check and spell out if not} domain, tools such as LegUp have implemented FPGA specific LLVM \gc{acronym first use, and no citation to LLVM?} passes to do hardware optimizations and scheduling~\cite{Canis2013LegUp:Systems}. In the software domain, LLVM implements an instruction combining pass as a set of rewriting rules. \gc{Not really sure what this paragraph is about?}

In this work we use a rewrite driven approach to the datapath optimization problem. The most relevant prior academic work is from Verma, Brisk and Ienne~\cite{dataflow2008verma}, who automatically apply dataflow transformations to combinational circuit designs. This work was inspired by the observation that ASIC logic synthesis tools could effectively deploy carry-save representation when presented with consecutive arithmetic operations, generating optimized netlists. However, when given arithmetic blocks separated by additional logical operations, the tool was not able to move the logical operations to facilitate optimal clustering. To address this issue, the authors designed specific logic arithmetic interchange rewrites that produced circuit designs, which when passed to logic synthesis could maximally cluster arithmetic operators together. By leveraging and extending these rewrites in the e-graph optimization framework, we can reproduce and extend results from this paper \cite{Verma2008VariableDesign}. \change{In addition to the work of Verma, Brisk and Ienne, a general purpose and verified RTL rewriting framework has been developed by Carl Seger and collaborators~\cite{Seger2023VossII,Pope2023Bifrost:Blocks}. The Voss II framework provides a design visualization environment and proposes a more interactive design space exploration approach, with little emphasis on automation. Datapath rewriting has also been applied to specific design challenges such as the design of large bitwidth multipliers~\cite{Ustun2022IMpress:HLS}. In datapath verification, rewriting has proven to be an invaluable technique~\cite{Koelbl2009SolverChecking,Coward2023DatapathRewriting,Yu2016AutomaticDatapath}.}

Although we target general datapath RTL optimization, there are several problem domains that have received particular attention and can be captured in the framework we present. One such instance is the multiple constant multiplication (MCM) problem~\cite{Gustafsson2007AProblems,Hartley1996SubexpressionMultipliers,Kumm2016MultipleArrays}, where the design problem is as follows: given a set of integer coefficients $\{a_1,...,a_{n}\}$, find an optimal circuit producing all the outputs $a_i \times x$ for a variable input $x$. The challenge presents many non-obvious operator sharing opportunities and is beyond the reach of existing logic synthesis tools. These problems are usually solved by hand or with bespoke tools, which represent constants using a fixed representation~\cite{Hartley1996SubexpressionMultipliers} such as Canonical Signed Digit (CSD)~\cite{Ercegovac2004DigitalArithmetic}. Alternative approaches deploy adder graph algorithms~\cite{Gustafsson2007AProblems} or have encoded the problem as an \change{integer linear programming problem~\cite{DeDinechin2021TowardsDesign,Kumm2018OptimalProgramming,Garcia2023TowardCost} or as Boolean satisfiability problem~\cite{Fiege2024Bit-LevelSatisfiability}.} Owing to the generality of the e-graph rewriting framework, such MCM optimizations are another class of methods that are automatically subsumed within our ROVER framework as a special case. \change{Note that bespoke MCM tools will outperform ROVER on complex MCM problems, as we shall see in Section~\ref{sect:results}.}
% \ssc{Some more general comment on datapath opto or rewriting?}

\begin{figure}
    \centering
    \subfloat[Initial e-graph contains \newline $(2\times x)>>1$\label{fig:egraph_stage_0}] {\includegraphics[scale=0.4]{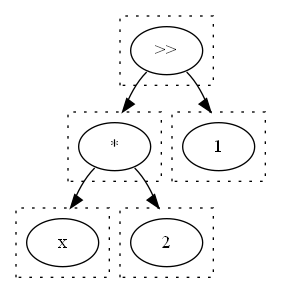}}
    % \qquad
    \subfloat[Apply $x\times 2 \rightarrow x<<1$\label{fig:egraph_stage_1}] {\includegraphics[scale=0.4]{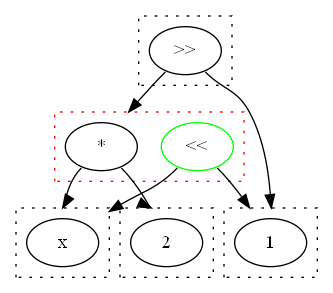}}
    \qquad
    \subfloat[Apply $(x<<s)>>s \rightarrow x$\label{fig:egraph_stage_2}] {\includegraphics[scale=0.4]{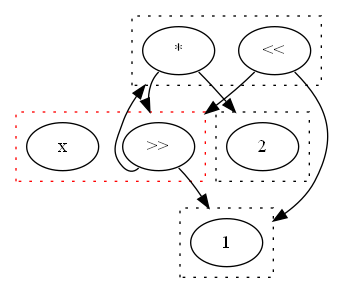}}
    \caption{E-graph rewriting for standard integer arithmetic. Dashed boxes represent e-classes of equivalent expressions. Green nodes represent newly added nodes. Red dashed boxes highlight which e-class has been modified.
    }
    \label{fig: e-graph_example}
\end{figure}
\subsection{E-Graphs}\label{subsec:egraphs}
An e-graph is a data structure developed in the theorem proving community~\cite{Nelson1980TechniquesVerification}. E-graphs provide a compact representation of equivalence classes (e-classes) of expressions. An e-graph represents a set of equivalent expressions, where nodes represent variables, constants or functions clustered together in e-classes. Edges represent operator inputs and connect nodes to e-classes, as shown in Figure~\ref{fig: e-graph_example}. In this way, a small number of nodes can represent exponentially many more expressions. Figure~\ref{fig: e-graph_example} contains a series of e-graphs, demonstrating how multiple equivalent expressions can be represented. 

E-graphs are grown using a technique called equality saturation~\cite{Willsey2021Egg:Saturation,Tate2009EqualityOptimization,Joshi2002Denali:Superoptimizer}, where rewrites that define equivalent expressions are applied to the e-graph. E-graphs are based on the theory of uninterpreted functions therefore an operator, e.g. addition, only gains any meaning via rewrites defined on that operator. For example $x+x \rightarrow 2 \times x$ tells the e-graph that $x+x$ is equivalent to $2\times x$. Given a set of rewrites, rewriting opportunities are detected via a process known as e-matching that identifies expressions in the e-graph that match the left-hand side patterns~\cite{DeMoura2007EfficientSolvers,Detlefs2005Simplify:Checking}. A key differentiating property of e-graphs is constructive rewrite application. \change{More formally, a rewrite is defined as a pair of expressions ($lhs$, $rhs$), such that when an expression syntactically matching the $lhs$ is discovered in the e-graph, the $rhs$ is added to the matched e-class in the e-graph.} The left-hand side is not destroyed and remains in the e-class. This means that the e-graph grows monotonically and application of one rewrite does not remove alternative rewriting opportunities. \change{Figure~\ref{fig: e-graph_example} shows an e-graph before (Figure~\ref{fig:egraph_stage_0}) and after (Figure~\ref{fig:egraph_stage_1}) application of a constructive rewrite that adds a new node to the e-graph. The final rewrite applied to produce Figure~\ref{fig:egraph_stage_2}, $(x\ll 1) \gg 1 \rightarrow x$, adds no new nodes to the e-graph, since the right-hand side expression is already contained within the e-graph. In this case, rewriting merges two existing equivalence classes, which, in this instance, leads to a loop in the e-graph.} 

Traditional rewrite engines suffer from the phase-ordering problem, which describes how the order of rewrite application can impact the final outcome~\cite{Kulkarni2012MitigatingLearning}. \change{Consider the following two ways to rewrite the initial expression from Figure~\ref{fig:egraph_stage_0}.}
\begin{align*}
    \text{Order 1: }&(x\times2)\gg 1 \rightarrow (x\ll 1) \gg 1 \rightarrow x \\
    \text{Order 2: }&(x\times2)\gg 1 \rightarrow (x+x) \gg 1
\end{align*}
\change{A destructive rewriting process leads to two different endpoints depending on the order of rewrite application, hence the phase-ordering problem. This problem is avoided entirely by constructive rewrite application~\cite{Willsey2021Egg:Saturation}. Applying the rewrite in ``Order 2'', the $x\times 2$ expression is retained and can be matched again, facilitating the application of ``Order 1''.} This will prove to be a particularly valuable property for hardware design. 

A general purpose and reusable e-graph library, {\egg}~\cite{Willsey2021Egg:Saturation}, was recently released and has fueled a new wave of e-graph research. In addition to its usability, {\egg} provides innovations in e-graph performance and numerous analysis features. In this work we will exploit the ability to write conditional and dynamic rewrite rules as well as the e-class analysis framework~\cite{Willsey2021Egg:Saturation}. The validity of conditional rules can be determined at runtime based on the specific values matched by the left-hand side pattern. The dynamic rewrite rules construct, at runtime, the right-hand side of a rewrite having matched a pattern. The e-class analysis feature allows users to attach additional information to an e-class, enabling program analysis techniques~\cite{Coward2022AbstractE-Graphs}. Since e-graphs grow monotonically, they usually reach a fixed point called saturation, where no further rewrite applications add additional information to the e-graph.

Proof production was recently added to \egg~allowing users to extract a rewrite sequence mapping one expression to an equivalent expression in the e-graph~\cite{Flatt2022SmallClosure}. This enables translation validation techniques to be applied to e-graph applications. Translation validation is a compiler technique to verify the correctness of a compiler's output~\cite{Cooper2011EngineeringEdition}. The verification problem is broken down into a sequence of sub-problems, verifying each step of the transformation. The proof production feature has been leveraged to develop an RTL verification assistant~\cite{Coward2023DatapathRewriting}.

E-graphs can be found in widely used SMT solvers such as Z3~\cite{DeMoura2008Z3:Solver}. More recently, {\egg} has helped to automate numerical stability improvement in the Herbie tool~\cite{Panchekha2015AutomaticallyExpressions} and synthesis smaller and more efficient rewrite sets via the Ruler tool~\cite{Nandi2020SynthesizingTransformations}. In the hardware domain, there is growing interest, with Ustun, Yu and Zhang advocating e-graph rewriting~\cite{Ustun2023EqualityOptimality}. Previous datapath research has explored alternative implementations of large multipliers on FPGAs, where different levels of decomposition were efficiently explored via equality saturation~\cite{Ustun2022IMpress:HLS}. ROVER tackles the more general ASIC RTL optimization problem, maximally exploiting logic synthesis capabilities. 

%%%%%%%%%%%%%%%%%%%%%%%%%%%%%%%%%%%%%%%%%%%%%%%%%%%%%%%%%%%%%%%%%
% LANGUAGE
%%%%%%%%%%%%%%%%%%%%%%%%%%%%%%%%%%%%%%%%%%%%%%%%%%%%%%%%%%%%%%%%%
\section{Intermediate Representation}\label{sect:language}

\begin{table}
    \centering
    \caption{VeriLang operators including the architecture used for theoretical cost assignment. Operators above the dashed line are those that directly translate from Verilog, whilst those below are custom operators that allow VeriLang to express more optimizations.}
    \begin{tabular}{lccl}
        Operator                         & Symbol & Arity & Architecture \\
        \midrule            
        Add/Sub             & +/-&  2                 & Prefix Adder (PA) \\ 
        Negation                         & - &  1                  & PA \cite{Beaumont-Smith2001ParallelDesign}\\
        Multiplication                   & $\times$                &  2    & Booth Radix-4 \cite{Koren2018ComputerAlgorithms}  \\
        % Gates                            & $\&,|,\oplus$           & 2 & Single Gate\\
        % Inverse Gates                    & $\sim\& ,\sim |, \sim\oplus$           & 2 & Single Gate\\
        Reduce                  & $\&,|,\hat{}$              & 1   & Log Tree\\
        Inverse Reduce          & $\sim\&,\sim|,\sim\hat{}$  & 1   & Log Tree\\
        Shifting                         & $\ll,\gg$               & 2   & Mux Tree \\
        Multiplexer                      & $\cdot ? \cdot : \cdot$ & 3   & Mux Gates \\
        Concat/Repl            & $\{,\}$                 & $n$ & Wiring \\
        \multirow{3}{*}{Comparison}                                 & $==,!=$             & &    \\
                               & $<,\leq$           & 2 & PA\\
                                         & $>,\geq$           &  & \\
        % Registers                        & \texttt{REGP}/\texttt{REGN} & 4 & Flip Flops\\
        \midrule
        Range Select                     & \texttt{slice} & 1 & Wiring\\
        Sum                              & \texttt{SUM} &  n    & CSA and PA\\
        Muxed Mult Array                        & \texttt{MUXAR}&  3    & Reduction and PA \\
        Fused Mult-Add               & \texttt{FMA} & 3 & Booth Radix-4 \\
        %Union (UNION)        & \\
    \end{tabular}
    \label{tab:operator_arch}
\end{table}

To facilitate RTL exploration via e-graph rewriting, we have developed an intermediate language, VeriLang, along with a parser and generator for translation to and from Verilog/SystemVerilog~\cite{Thomas2008TheLanguage}. Since e-graphs work with expressions, VeriLang is a nested S-expression language in Common Lisp~\cite{Steele1990CommonLanguage}. A formal description is given in Grammar~\ref{gr:verilang_grammar}. 
% \gc{Finally, I think your caption might need a bit more detail, e.g. you can say that {\em var} is a terminal variable symbol drawn from some defined set of expression variables.}
\begin{grammar}[VeriLang grammar definition. The terminal variable $var$ is a symbol drawn from a set of expression variables, and $op$ is an operation from the supported set of VeriLang operators as described in Table~\ref{tab:operator_arch}.][h][gr:verilang_grammar]
\firstcase{\texttt{term}}{(op\;\; \texttt{width}\;\; [\texttt{arg}]\;$\ldots$\; [\texttt{arg}])}{}
\otherform{var \gralt \texttt{int}}{}
\firstcase{\texttt{\texttt{arg}}}{\texttt{width}\;\;\texttt{signage}\;\;\texttt{term}}{}
\firstcase{\texttt{\texttt{width}}}{var \gralt \texttt{int}}{}
\firstcase{\texttt{\texttt{signage}}}{var \gralt \texttt{unsign} \gralt \texttt{sign}}{}
\end{grammar}

As an example, in VeriLang, an 8-bit unsigned addition, stored in a 9-bit result would be expressed as:
\begin{equation} \label{eqn:add_verilang}
\texttt{(+ 9 8 unsign x 8 unsign y)}.    
\end{equation}
To provide VeriLang semantics, we first specify two functions.
\begin{align}
    \llbracket \cdot \rrbracket:& \texttt{term} \rightarrow \mathbb{Z}\\
    \cdot_{\cdot,\cdot} :& \mathbb{Z}\times \mathbb{N} \times \{\texttt{unsign}, \texttt{sign}\} \rightarrow \mathbb{Z}
\end{align}
We then define the semantics in terms of integer arithmetic:
\begin{align}
    &\left \llbracket (op\; w\; w_1\;s_1\;t_1\;\ldots\; w_n\;s_n\;t_n)\right \rrbracket =\\
    &\left (\llbracket op \rrbracket\; \llbracket t_1\rrbracket_{w_1,s_1}\;\ldots\; \llbracket t_n\rrbracket_{w_n,s_n}\right)_{w,\texttt{unsign}}
\end{align}
where $\llbracket op \rrbracket$ denotes the standard interpretation of $op$ acting on integers and for $k \in \mathbb{Z}$, $w \in \mathbb{N}$ and $s\in \{\texttt{unsign}, \texttt{sign}\}$,
\begin{equation}
    k_{w,s} = \begin{cases}
    k \mod{2^w}, \hspace{1.3cm}\text{if } s == \texttt{unsign}\\
    2(k \mod{2^{w-1}})-(k \mod{2^w}),              \text{else.}
\end{cases} 
\end{equation}
This is a valid model of bitvector arithmetic under the least positive residue definition of modulus.
\begin{equation}
    \cdot \mod{\cdot}: \mathbb{Z}\times \mathbb{N} \rightarrow \mathbb{N}
\end{equation}
Under these semantics, \eqref{eqn:add_verilang} has the following interpretation:
\[
\left(+ \hspace{1.5em}\left(\llbracket x \rrbracket \mod{2^8}\right)\hspace{1.5em}\left(\llbracket y \rrbracket \mod{2^8}\right)\right) \mod{2^9}.
\]

% \gc{We don't have a section on the semantics of the operators, which I guess is fine, but the above example I think needs at least a passing reference to the assumed semantics for it to make sense? i.e. here `8 unsign b' means `take b as an arbitrary width integer, truncate the bit-level representation to an 8-bit integer, and interpret the resulting value as an unsigned integer. We could define this formally -- in fact, perhaps we should on the whiteboard, whether you integrate into this paper or not. Without this, it's a bit implicit what's going on here and why this makes sense.}

% Here, the addition operator takes eight operands, of which six are type annotations denoting input and output bitwidth and signage interpretations. For example, \texttt{8 unsign y} means take \texttt{y} as an arbitrary width integer and truncate the most-significant bits of the bit-level representation to an 8-bit integer, then interpret the bit vector of length 8 as an unsigned integer. The \texttt{9 unsign} means that the output of the addition operation is stored in a 9-bit unsigned integer. 

% These type arguments could more intuitively be thought of as edge labels. The remaining two operands are the variables being summed. 
Type annotations are essential, since Verilog is a context determined language. The signage of an operator is determined by the signage of its input operands. For this reason we do not include a signage annotation for the output of an operator in VeriLang. The bitwidth of an operator is determined by the bitwidth of the largest operand, {\em including the left-hand side of an assignment}~\cite{Thomas2008TheLanguage}. Therefore we do include a bitwidth annotation for the output of an operator in VeriLang. Only the subset of VeriLang expressions comprised of concrete instances of the \texttt{width} and \texttt{signage} type parameters, meaning these cannot be variables, can be translated to synthesizable Verilog. 

% The first component of ROVER is an intermediate language (IL) definition that can express RTL at the abstraction level of System Verilog operating on fixed width bitvectors \textbf{reference system verilog}. The IL also captures signage, where a bitvector can be interpreted in 2's complement representation or as an unsigned bitvector. Capturing both bitwidth and signage in the IL is essential as the behaviour of operators in Verilog is defined by both of these properties of it's operands. 
Since e-graph rewriting is based on the theory of uninterpreted functions, operators take on meaning via rewrites that define equivalent implementations. VeriLang is designed with rewrites in mind, making it simple to express conditional and dynamic rewrites with access to all the relevant parameter values. In Section~\ref{sect:rewrites} we describe how ROVER's rewrites differentiate between type annotations and variables. Type annotations are also essential for accurate hardware costing, since an 8-bit addition should be cheaper than a 32-bit addition. 

VeriLang currently supports almost all the fundamental Verilog operators, with the exception of less commonly used operators such as trigger (\texttt{->}), modulus (\%) and power (**), though these could easily be added. In total we support 29 of the Verilog defined operators as shown in Table~\ref{tab:operator_arch}, which omits the single gate operators that are also supported. 

%Registers are represented as \texttt{REGP} and \texttt{REGN} operators, that allow ROVER to represent pipelined circuit designs.  \gc{Not obvious what the difference is between REGP and REGN.... I'm guessing it's initialisation?? But that would only make sense for single-bit registers, right? Maybe you can explain to me and also ensure this question is not left hanging in the manuscript? SAM - this is because one updates on the posedge and one updates on the negedge}
% We additionally include a \texttt{slice} operator that enocodes bit selection of a bitvector.

\begin{figure}
    \centering
    % \subfloat[Consecutive additions]{\includegraphics[width=.35\columnwidth]{two_adds.png}}
    \subfloat[Consecutive additions]{\includegraphics[scale=.425]{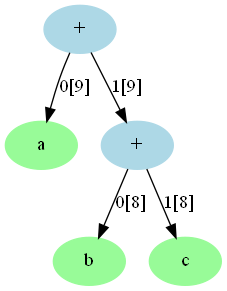}}
    \qquad
    % \subfloat[Merged additions encoded as a \texttt{SUM}] {\includegraphics[width=.55\columnwidth]{two_sum.png}
    \subfloat[Merged additions encoded as a \texttt{SUM}] {\includegraphics[scale=.425]{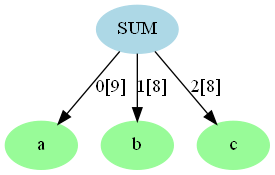}
    \label{fig: merged_adds}}
    \caption{Edge labels show the operand's index and bitwidth in square brackets.}
    \label{fig: sum_node}
\end{figure}

In addition to the Verilog operators, VeriLang supports a set of custom operators as described in Table \ref{tab:operator_arch}, which capture the optimization capabilities of modern ASIC logic synthesis tools. These additional operators greatly improve correlation between ROVER's cost model and the final circuit cost reported by commercial synthesis tools~\cite{Zimmermann2009DatapathDesign}. The \texttt{SUM} operator encodes how multiple additions can be clustered into a single carry-save adder (CSA) allowing the circuit to deploy fewer expensive carry-propagate adders. These clustering nodes are typically valuable but may not be useful if an intermediate result is required. Figure~\ref{fig: sum_node} shows two consecutive additions being reduced to a single \texttt{SUM} node. We include two further merged operators, the familiar Fused Multiply-Add (FMA), which encodes the ability to construct the circuit for $a*b + c$ using a single carry-propagate adder and the Muxed Mult Array, which encodes the synthesis optimization for $a\times b + \bar{a}\times c$ as described in~\cite{Zimmermann2009DatapathDesign}. The Muxed Mult Array will be discussed further in Section~\ref{sect:rewrites}.

As shown in Figure~\ref{fig:flow_diagram}, the input Verilog/System Verilog is first parsed by the open-source \texttt{slang} parser~\cite{Popoloski2023Slang}, generating a JSON representation. The ROVER front-end then translates this JSON representation into a VeriLang expression. From this VeriLang expression \egg~generates an initial e-graph, where each e-class contains a single node. In the initial translation phase, we construct a mapping from the original variable names to their corresponding VeriLang expressions. By attaching the variable name to the corresponding e-classes, we retain information from the original RTL, which we can use during code generation (Section~\ref{sec:code_gen}) to improve readability.

%%%%%%%%%%%%%%%%%%%%%%%%%%%%%%%%%%%%%%%%%%%%%%%%%%%%%%%%%%%%%%%%%
% REWRITES
%%%%%%%%%%%%%%%%%%%%%%%%%%%%%%%%%%%%%%%%%%%%%%%%%%%%%%%%%%%%%%%%%
\section{Rewrites}\label{sect:rewrites}

\subsection{Specifying Rewrites} \label{subsect:spec_rewrites}
Rewrites define local equivalences between two expressions that, when chained, enable architectural exploration. Equivalence is defined as functional equivalence over \texttt{term}s in VeriLang. Namely, given \texttt{term}s $t_1$ and $t_2$, $t_1\cong t_2$ if and only if for all possible inputs $t_1$ and $t_2$ produce identical outputs under the semantics of VeriLang. A rewrite is defined as a transformation from a \texttt{term} to a \texttt{term}. Note that rewrite pattern \texttt{term}s may contain free variable bitwidth and signage type parameters. This is analogous to using parameterizable bitwidths in Verilog as opposed to concrete integer values. 

Via the e-matching process described in Section~\ref{subsec:egraphs}, \egg~matches a \texttt{term} in the e-graph returning a \texttt{map} that is an assignment of (some of the) variables in the \texttt{term} to concrete values. A partial evaluation of a \texttt{term} with respect to a \texttt{map} produces a new \texttt{term}, $\llbracket \cdot \rrbracket_\cdot : \texttt{term} \times \texttt{map}\rightarrow \texttt{term}$. As a first example, we describe the unconditional commutativity rewrite that is always valid. Later we will give an example of a conditional rewrite. Commutativity of addition is defined as:
\begin{equation*}
    \overbrace{(+\;w\;w_a\,s_a\,a\;w_b\,s_b\,b)}^{lhs} \rightarrow
    \overbrace{(+\;w\;w_b\,s_b\,b\;w_a\,s_a\,a)}^{rhs}.
\end{equation*}
    
If applied to an e-graph containing \eqref{eqn:add_verilang}, the e-matching process would return a \texttt{map}, 
%% GC addition starts
\begin{equation}
m = \left\{ \begin{array}{ll} 
w  & \mapsto 9 \\ 
w_a& \mapsto 8 \\ 
s_a& \mapsto \texttt{unsign} \\ 
w_b& \mapsto 8 \\
s_b& \mapsto \texttt{unsign} 
 \end{array} \right.
\end{equation}
%% GC addition ends
Note that $m$ is a partial function because it does not provide any assignment for variables $a$ and $b$. This approach differs from other e-graph based applications, in that a single rewrite encodes a rewrite over many distinct types. Previous work encoded types in the operator name itself e.g. $+_{16}$ and $\times_{32}$~\cite{Ustun2022IMpress:HLS}, but in our setting this is impractical due to the number of operators we would have to support. The partially evaluated \texttt{term}, $\llbracket rhs \rrbracket_m$ is then added to the e-graph, where
\begin{equation*}
\llbracket rhs \rrbracket_m = \texttt{(+ 9 8 unsign b 8 unsign a)}.
\end{equation*}
For this simple commutativity example, the rewrite is valid anywhere that it matches. However, the set of RTL rewrites for which this statement holds is small. To enable meaningful RTL transformations, we define a set of conditionally applied rewrites specified as a triple $(\texttt{cond}, \texttt{term}, \texttt{term})$, where
\begin{equation*}
\texttt{cond}: \texttt{map} \rightarrow \texttt{Bool}.
\end{equation*}
The condition is checked each time the left-hand side \texttt{term} of a rewrite is matched. The partially evaluated right-hand side is only added to the e-graph if the condition returns true. That is, the condition for correctness of a conditional rewrite  $(\phi, lhs, rhs)$ is that for any map $m$:
\begin{equation}\label{eqn:cond_rewrite}
    \phi(m) \Rightarrow \llbracket lhs \rrbracket_m \cong \llbracket rhs \rrbracket_m.
\end{equation}

In Figure~\ref{fig:verilog_rewriting}, we provide an example, to highlight where the validity of a rewrite can depend on the context. Specifically, the associativity rewrite is valid in the case where the intermediate signal retains the carry-out of the first addition. 

Conditional rewriting allows ROVER to detect all syntactic opportunities to apply a transformation and then filter out those that would be semantically invalid. Such an approach allows ROVER to capture a wide range of RTL transformations without sacrificing correctness. In Section~\ref{subsect:rewrite_validation} we describe the construction of the conditions and return to this example to construct a condition for this exact associativity rewrite.

\begin{figure}
    \centering
\begin{lstlisting}[language=Verilog, basicstyle=\footnotesize]
wire  [7:0] A, B, C;
wire  [7:0] add_8bit;
wire  [8:0] add_9bit, add_right;
wire  [9:0] left1, left2, right;
    
assign add_8bit  = A + B; // carry-out discarded
assign left1     = add_8bit + C; 

assign add_9bit  = A + B; // carry-out retained
assign left2     = add_9bit + C; 

assign add_right = B + C;
assign right     = A + add_right; 
\end{lstlisting}
    \caption{Verilog associativity rewriting example. Signals \texttt{left1} and \texttt{right} are functionally distinct, because the carry-out is discarded in computing \texttt{add\_8bit}, therefore \texttt{left1}$\not\rightarrow$\texttt{right}. The signals \texttt{left2} and \texttt{right} are functionally equivalently, therefore it is valid to rewrite \texttt{left2}$\rightarrow$\texttt{right}.}
    \label{fig:verilog_rewriting}
\end{figure}
% \gc{I think the conditionality point is important to the novelty of this work, but doesn't come across quite as strongly as I'd like,} 
%partly because the example of a condition is not given for Fig 5, just the example -- not sure if that's an Intel IP restriction, or not. Another simple way to emphasise this is that when you first introduce your unconditional commutativity rewrite above, explicitly say it's a first simple example of an unconditional rewrite -- before you even introduce it -- and that you'll also give an example of a conditional rewrite later on.}
The set of rewrites described in Table \ref{tab:rewrites} captures optimizations learnt from Intel's Numerical Hardware Group, prior work~\cite{dataflow2008verma} and logic synthesis documentation~\cite{Synopsys2019CodingSynthesis,Zimmermann2009DatapathDesign}. All rewrites include the type annotations described in Section~\ref{sect:language}. We impose no restrictions on the bitwidth and signage parameters in the rewrites, to ensure maximum generality of the rewrites. We omit the bitwidth and signage annotations as well as the conditions in Table~\ref{tab:rewrites} to maintain readability. 

%We shall denote rewrites $l(\vec{p})\rightarrow r(\vec{p})$, where $\vec{p}$ is a set of type annotations for the inputs and outputs and $l,r$ are functions mapping type annotations to particular expressions. 

% For each rewrite, ROVER checks an internal condition to ensure validity of the particular instance. The conditions are functions of the type annotations and are discussed in Section~\ref{subsect:rewrite_validation}. 
% and not described the conditions under which the rewrites are valid unless very simple. Further details on condition generation are described in Section \ref{subsect:rewrite_validation}, but most of the conditions are Boolean expressions of 10's or 100's of terms. 

\begin{table*}
    \centering
    \caption{ROVER's bitwidth and signage dependent datapath rewrites. Bitwidth and signage parameters are omitted here. The $*$ operation represents both $\{+,\times\}$. The rules are conditionally applied as a function of the bitwidth and signage information attached to each operand. The necessary and sufficient conditions are too complex (denoted by $\dagger$) to display in column 4 for most rewrites.}
    \begin{tabular}{ccccc}
    \toprule
    
    Class & Name & Left-hand Side & Right-hand Side & Condition \\
    \midrule
    
    \multirow{12}{*}{Bitvector Arithmetic} 
    & Commutativity                  & $ a * b$ 
                                     & $ b * a$ 
                                     & True \\
    %----------------------------------------------------------------------------------------------------------------------------------------------
    & \mycc Associativity            & \mycc  $(a * b) * c$ 
                                     & \mycc  $a * (b * c)$ 
                                     & \mycc  $ \dagger $ \\
    %----------------------------------------------------------------------------------------------------------------------------------------------
    & Associativity of Sub           & $(a - b) - c$ 
                                     & $a - (b + c)$ 
                                     & $ \dagger $ \\
    %----------------------------------------------------------------------------------------------------------------------------------------------
    & \mycc Dist Mult over Add/Sub  &\mycc  $a \times (b \pm c) $
                                    &\mycc $ (a \times b) \pm (a \times c)$ 
                                    &\mycc $ \dagger $ \\
    %----------------------------------------------------------------------------------------------------------------------------------------------
    & Dist Add/Sub over Mult       & $ (a \times b) \pm (a \times c)$ 
                                   & $ a \times (b \pm c) $ 
                                   & $ \dagger $ \\
    %----------------------------------------------------------------------------------------------------------------------------------------------
    &\mycc Add Zero                  &\mycc $a + 0$
                                     &\mycc $\texttt{slice}(a)$ 
                                     &\mycc $ \dagger $ \\
    %----------------------------------------------------------------------------------------------------------------------------------------------
    & Mul by Zero                    & $a \times 0$
                                     & $0$ 
                                     & $ \dagger $ \\
    %----------------------------------------------------------------------------------------------------------------------------------------------
    &\mycc  Mult by One              &\mycc $a \times 1$
                                     &\mycc $\texttt{slice}(a)$ 
                                     &\mycc True \\
    %----------------------------------------------------------------------------------------------------------------------------------------------
    &  Mult by Two                   & $a \times 2$
                                     & $a \ll 1$ 
                                     & True \\      
    %----------------------------------------------------------------------------------------------------------------------------------------------
    &\mycc  Sub to Neg               &\mycc  $a - b$
                                     &\mycc $a + (-b)$ 
                                     &\mycc True \\
    %----------------------------------------------------------------------------------------------------------------------------------------------
    &  Sum Same                      & $ a + a $
                                     & $ 2 \times a$
                                     & $ \dagger $ \\
    %----------------------------------------------------------------------------------------------------------------------------------------------
    &\mycc  Mult Sum Same            &\mycc $ (a \times b) + b$
                                     &\mycc $ (a + 1) \times b$ 
                                     &\mycc $ \dagger $ \\
    \hline
    %----------------------------------------------------------------------------------------------------------------------------------------------
    %----------------------------------------------------------------------------------------------------------------------------------------------
    \multirow{8}{*}{Bitvector Logic} 
    & Merge Left Shift        & $ (a\ll b)\ll c$
                              & $  a \ll (b+c)$
                              & $ \dagger $ \\
    %----------------------------------------------------------------------------------------------------------------------------------------------
    &\mycc Merge Right Shift  &\mycc $(a\gg b) \gg c$
                              &\mycc $ a \gg (b+c)$  
                              &\mycc $ \dagger $ \\
    %----------------------------------------------------------------------------------------------------------------------------------------------
    & Redundant Sel           & $ b ? a : a $
                              & $\texttt{slice}(a)$ 
                              & True \\
    %NEW----------------------------------------------------------------------------------------------------------------------------------------------
    &\mycc Nested Mux Left   &\mycc $ a\,?\,(a\,?\, b : c) : d$
                             &\mycc $ a\,?\,b : d$ 
                             &\mycc $ \dagger $ \\
    %NEW----------------------------------------------------------------------------------------------------------------------------------------------
    & Nested Mux Right       & $a\,?\,b : (a\,?\, c : d)$
                             & $a\,?\,b : d$ 
                             & $ \dagger $ \\
    %NEW----------------------------------------------------------------------------------------------------------------------------------------------
    &\mycc Sel Left Shift    &\mycc  $ e ? (a \ll b) : (c \ll d)$
                             &\mycc  $(e ? a : c) \ll (e ? b : d)$ 
                             &\mycc $ \dagger $ \\
    %NEW----------------------------------------------------------------------------------------------------------------------------------------------
    & Sel Right Shift        &  $ e ? (a \gg b) : (c \gg d)$
                             &  $(e ? a : c) \gg (e ? b : d)$ 
                             & $ \dagger $ \\
    %----------------------------------------------------------------------------------------------------------------------------------------------
    &\mycc Not over Con       &\mycc $\sim\{a,b\}$
                              &\mycc $\{(\sim a),(\sim b)\}$ 
                              &\mycc $ \dagger $ \\
    \hline
    %----------------------------------------------------------------------------------------------------------------------------------------------
    %----------------------------------------------------------------------------------------------------------------------------------------------
    \multirow{11}{7em}{\centering Arithmetic Logic Exchange} 
    & Left Shift Add         & $ (a + b) \ll c $
                                   & $ (a \ll c) + (b \ll c)$ 
                                   & $ \dagger $ \\
    %----------------------------------------------------------------------------------------------------------------------------------------------
    &\mycc Add Right Shift         &\mycc  $a + (b \gg c)$
                                   &\mycc  $((a \ll c) + b) \gg c$ 
                                   &\mycc  $ \dagger $ \\
    %----------------------------------------------------------------------------------------------------------------------------------------------
    & Left Shift Mult              & $(a\times b)\ll c$
                                   & $(a\ll c)\times b$ 
                                   & $ \dagger $ \\
    %----------------------------------------------------------------------------------------------------------------------------------------------
    & \mycc Sel Add/Mul            &\mycc  $ e ? (a * b) : (c * d)$
                                   &\mycc  $(e ? a : c) * (e ? b : d)$ 
                                   &\mycc $ \dagger $ \\
    
    %----------------------------------------------------------------------------------------------------------------------------------------------
    &  Sel Add Zero Left           & $e ? (a + b) : c$
                                   & $(e ? a : c) + (e ? b :  0)$ 
                                   & $ \dagger $ \\
    %----------------------------------------------------------------------------------------------------------------------------------------------
    &\mycc  Sel Add Zero Right     &\mycc $ e ? a : (b + c)$
                                   &\mycc $(e ? a : b) + (e ? 1 :  c)$ 
                                   &\mycc $ \dagger $ \\
    %NEW-ish----------------------------------------------------------------------------------------------------------------------------------------------
    &  Sel Mul One Left            & $e ? (a \times b) : c$
                                   & $(e ? a : c) \times (e ? b :  1)$ 
                                   & $ \dagger $ \\
    %NEW-ish----------------------------------------------------------------------------------------------------------------------------------------------
    &\mycc  Sel Mul One Right      &\mycc $ e ? a : (b \times c)$
                                   &\mycc $(e ? a : b) \times (e ? 1 :  c)$ 
                                   &\mycc $ \dagger $ \\
    %----------------------------------------------------------------------------------------------------------------------------------------------
    &  Move Sel Zero               & $ (b ? 0 : a) \times c$
                                   & $ a \times (b ? 0 : c)$ 
                                   & $ \dagger $ \\
    %----------------------------------------------------------------------------------------------------------------------------------------------
    &\mycc  Concat to Add          &\mycc $\{a,b\}$
                                   &\mycc $(a\ll w_b) + b$ 
                                   &\mycc $ \dagger $ \\
    %----------------------------------------------------------------------------------------------------------------------------------------------
    &  Neg Not                     & $- a$
                                   & $(\sim a) + 1$ 
                                   & $ \dagger $ \\
    \hline
    %----------------------------------------------------------------------------------------------------------------------------------------------
    %----------------------------------------------------------------------------------------------------------------------------------------------
    \multirow{3}{*}{Merging Ops}
    & \mycc Merge Additions &\mycc $a1 + (a2 + (a3+...+an)...)$
                            &\mycc $\texttt{SUM}(a1, a2,..., an)$ 
                            &\mycc $ \dagger $ \\
    %----------------------------------------------------------------------------------------------------------------------------------------------
    & Merge Mult Array      & $(a \times b) + (c \times (\sim b))$  
                            & $\texttt{MUXAR}(b, a, c)$ 
                            & $ \dagger $ \\
    %----------------------------------------------------------------------------------------------------------------------------------------------
    &\mycc  FMA Merge       &\mycc $(a \times b) + c$ 
                            &\mycc $\texttt{FMA}(a, b, c)$ 
                            &\mycc $ \dagger $ \\
    \hline
    %----------------------------------------------------------------------------------------------------------------------------------------------
    %----------------------------------------------------------------------------------------------------------------------------------------------
    
    \multirow{2}{*}{Constant Expansion} 
    &  Mult Constant &  $ c \times x$
                          &  $((2\times (c \gg 1)) \times x) + (c[0] \times x)$ 
                          &  $ \dagger $ \\
    %----------------------------------------------------------------------------------------------------------------------------------------------
    & \mycc One to Two Mult     & \mycc $ 1\times x$
                          & \mycc $ (2\times x) - x$ 
                          & \mycc $ \dagger $ \\
    %----------------------------------------------------------------------------------------------------------------------------------------------
    \bottomrule
    \end{tabular}
    \label{tab:rewrites}
\end{table*}

ROVER combines both static rewrites, where the right-hand side is known at compile time, and dynamic rewrites, where the right-hand side is constructed at runtime. Dynamic rewrites are particularly useful for constant manipulation, building normal forms and computing sufficient bitwidths. 

The first group, bitvector arithmetic identities, contains familiar arithmetic rewrites allowing ROVER to re-arrange and simplify arithmetic expressions. The second group includes transformations more commonly encountered in hardware design, simplifying logical expressions and removing redundant logic. The third class of rewrites, Arithmetic Logic Exchange, are inspired by the work of Verma {\em et al.}~\cite{dataflow2008verma} and facilitate the discovery of additional arithmetic clustering opportunities. These opportunities can be missed by logic synthesis as arithmetic operations can be separated by logical operations. The Arithmetic Logic Exchange rewrites allow ROVER to move logic operations over arithmetic operations, enabling larger arithmetic clusters to form. Once clustered together, these blocks can be effectively optimized by logic synthesis resulting in more optimal circuit designs. We extend prior work on this subject~\cite{dataflow2008verma}, generalizing and expanding the scope.

The Merging Ops rewrites detect certain operator combinations and cluster them into a single custom operator which, as described in Section \ref{sect:language}, allows ROVER to identify sub-circuits that synthesis tools will specifically optimize~\cite{Synopsys2019CodingSynthesis}. Both the ``Merge Additions'' and ``FMA Merge'' rewrites exploit carry-save format to construct a multi-row array which can be reduced using half- and full-adders~\cite{Ercegovac2004DigitalArithmetic}. Like the \texttt{SUM} operator, the \texttt{FMA} operator requires a single carry-propagate adder to generate the result $a\times b + c$. The ``Merge Mult Array'' identifies disjoint multiplier arrays that can be merged. Letting $a[i]$ represent bit $i$ of $a$ and $u = \lceil\log_2 r \rceil$, \texttt{MUXAR} in the table denotes the right hand side of the rewrite, where the \texttt{SUM} represents array reduction:
\begin{align*} %\label{eqn:mux_array}
%(a &\times b) + (c \times (\sim b))) \rightarrow \\ % GC: deleted this
\texttt{MUXAR}(b,a,c) = \\  % GC: inserted this
\texttt{SUM}(&(b[0] ? a : c) \ll 0, \\
& (b[1] ? a : c)\ll 1,..., \\
& (b[r-1] ? a : c)\ll(r-1)).
\end{align*}
These rewrites help ROVER to identify the best design to pass onto logic synthesis as they encode downstream logic synthesis optimizations directly in the e-graph.

The remaining class of rewrites, ``Constant Expansion'', explores alternative representations of constants in hardware with particular attention paid to multiplication of a variable by a constant. These rules generalize MCM optimizations and are valuable where constant manipulation can occur as a sub-problem in a larger design optimization, where a specialist MCM tool is not applicable. We shall encounter such results in Section~\ref{sect:results}\change{, but will also encounter limitations of a rewriting approach for complex MCM problems}. These rules allow ROVER to re-create and generalise results from the MCM literature described in Section~\ref{subsect:datapath_opto}. As in previous \egg~implementations, constant folding is implemented as an e-class analysis~\cite{Willsey2021Egg:Saturation}. 

% Several rewrites use the $\texttt{slice}$ operator, for example ``Add Zero'', which is necessary because bitwidth and signage information is associated with an operator, not the operand itself. Converting an operator to a single operand would remove information in the representation, creating opportunities for invalid rewrites. The \texttt{slice} operators can be removed once the type information is transferred to its parent operator. We do not define any rewrites over registers, which means that ROVER does not modify the pipeline structure of an input design. \gc{Not sure this paragraph is required. The first part seems to be quite `in the weeds'. The part about pipelining seems superfluous -- if you don't define rewrites over registers, why are we bothering to talk about them? Seems odd? Let's discuss this.}

%%%%%%%%%%%%%%%%%%%%%%%%%%%%%%%%%%%%%%%%%%%%%%%%%%%%%%%%%%%%%%%%%
% CONDITIONAL REWRITES
%%%%%%%%%%%%%%%%%%%%%%%%%%%%%%%%%%%%%%%%%%%%%%%%%%%%%%%%%%%%%%%%%
\subsection{Synthesizing Rewrite Conditions} \label{subsect:rewrite_validation}

% \gc{I think the approach to the problem could be clarified here. I think this section is a bit confusing, in particular in terms of (1) the relationship between $T$ and $\phi$, (2) the domain over which validity of $\phi$ is guaranteed, (3) the algorithm for constructing $\phi$ from $T$. Let's discuss face-to-face the best way to explain all this.}

As described above, rewrites are encoded as triples $(\texttt{cond},\texttt{term}, \texttt{term})$, where the \texttt{term}s may contain variable \texttt{width} and \texttt{signage} parameters. Not all assignments to these parameters produce valid rewrites. Namely, in general, for rows in the table with $\dagger$ conditions, there exist mappings $m$ such that $\llbracket lhs \rrbracket_\texttt{m} \not\cong \llbracket rhs \rrbracket_m$. In this section we describe a solution to the following problem. Given a pair of \texttt{term}s, $(lhs,rhs)$, construct a \texttt{cond}, $\phi$, such that for all maps $m$,
\begin{equation}\label{eqn:necc_and_suff}
    \phi(m) \Leftrightarrow \llbracket lhs \rrbracket_m \cong \llbracket rhs \rrbracket_m.
\end{equation}
The sufficiency of $\phi$ $(\Rightarrow)$ is essential because applying a single invalid rewrite introduces a non-equivalent expression into the e-graph, meaning that no design in the e-graph can be trusted. The necessity of $\phi$ $(\Leftarrow)$ ensures that no rewriting opportunities are missed by ROVER. In practice, constructing a $\phi$ satisfying \eqref{eqn:necc_and_suff} is challenging. To make progress, we make certain assumptions that simplify the problem, as described below.

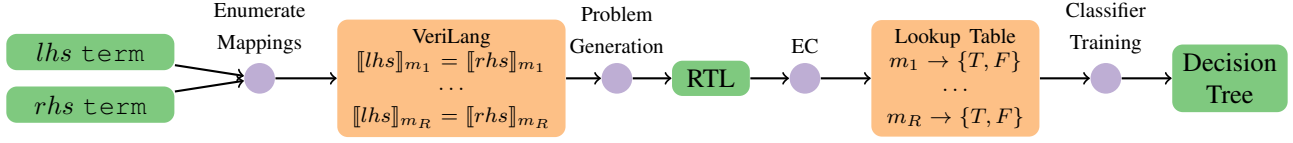
\begin{figure*}
    \centering
    \begin{tikzpicture}

% Boxes
\node [shape=rectangle, fill=scgreen, text width = 2cm, text centered,rounded corners] at (-9,0.35) (lhs) {$lhs$ \texttt{term}};
\node [shape=rectangle, fill=scgreen, text width = 2cm, text centered,rounded corners] at (-9,-0.35) (rhs) {$rhs$ \texttt{term}};
\node [shape=circle, fill=scpurple, text width = 0.1cm, text centered,label={[text width =2cm,text centered]\footnotesize{Enumerate Mappings}}] at (-6.75,0) (enum) {};
\node [shape=rectangle, fill=scorange, text width = 2.8cm, text centered,rounded corners] at (-4.2,0) (verilang_prob) {\footnotesize{VeriLang\\ $\llbracket lhs\rrbracket_{m_1} = \llbracket rhs\rrbracket_{m_1}$\\$\ldots$\\ $\llbracket lhs\rrbracket_{m_R} = \llbracket rhs\rrbracket_{m_R}$}};
\node [shape=circle, fill=scpurple, text width = 0.1cm, text centered,label={[text width =2cm,text centered]\footnotesize{Problem Generation}}] at (-2,0) (prob_gen) {};
\node [shape=rectangle, fill=scgreen, text width = 0.8cm, text centered,rounded corners] at (-0.75,0) (rtl) {RTL};

\node [shape=circle, fill=scpurple, text width = 0.1cm, text centered,label={\footnotesize{EC}}] at (0.5,0) (dpv) {};
% \node [shape=rectangle, fill=scorange, text width = 1.3cm, text centered, rounded corners] at (2,0) (lut) {Lookup Table};
\node [shape=rectangle, fill=scorange, text width = 2cm, text centered,rounded corners] at (2.5,0) (lut) {\footnotesize{Lookup Table\\ $m_1 \rightarrow \{T,F\}$\\$\ldots$\\ $m_R \rightarrow \{T,F\}$}};

\node [shape=circle, fill=scpurple, text width = 0.1cm, text centered,label={[text width=1.8cm, text centered]\footnotesize{Classifier Training}}] at (4.5,0) (classifier) {};
\node [shape=rectangle, fill=scgreen, text width = 1.3cm, text centered,rounded corners] at (6.15,0) (dec_tree) {Decision Tree};
% \node [shape=circle, fill=scpurple, text width = 0.1cm, text centered,label={[label distance=0.2cm]\footnotesize{ROVER Back End}}] at (7,0) (back_end) {};
% \node [shape=rectangle, fill=scgreen, text width = 2cm, text centered, rounded corners] at (9,0) (output) {Opt. Verilog};

\draw [->,thick] (lhs) edge (enum);
\draw [->,thick] (rhs) edge (enum);
\draw [->, thick] (enum) edge (verilang_prob);
\draw [->, thick] (verilang_prob) edge (prob_gen);
\draw [->, thick] (prob_gen) edge (rtl);
\draw [->, thick] (rtl) edge (dpv);
\draw [->, thick] (dpv) edge (lut);
\draw [->, thick] (lut) edge (classifier);
\draw [->, thick] (classifier) edge (dec_tree);

\end{tikzpicture}
    \caption{Flow diagram for the automated process of synthesizing rewrite conditions. The output is a decision tree that is translated into a Boolean expression.}
    \label{fig:rewrite_cond_flow_diagram}
\end{figure*}

We have developed an automated condition synthesis flow, shown in Figure~\ref{fig:rewrite_cond_flow_diagram}, that makes ROVER extensible. Developers or design engineers can specify new ROVER rewrite rules as pairs of \texttt{term}s and run ROVER's condition synthesis flow to automatically generate a correct \texttt{cond}. This allows design engineers to include valuable transformations drawing from their own experience, but avoids the overhead of considering all the scenarios in which the transformation is valid or invalid. The idea is to sample the space of all signages and all small bitwidth combinations, and to build a general rule for validity consistent with the sample taken.

The automated condition synthesis flow deploys program synthesis~\cite{Solar-Lezama2009ProgramSketching}, where a correct condition is learnt from data.
Let $lhs$ contain $H$ free bitwidth parameters $w_1$ to $w_H$ and $G$ free signage parameters $s_1$ to $s_G$.
% \begin{align*}
%     M = \{ &
%     \texttt{w1} \mapsto w_1, \ldots \texttt{wH} \mapsto w_H, \texttt{s1} \mapsto s_1, \ldots \texttt{sG} \mapsto s_G \\
%     &| w_i \in \{1,\ldots,8\} \wedge s_i \in \{\texttt{unsign},\texttt{sign}\}  \}.
% \end{align*}
% %% GC addition ends
\begin{align*}
    M = \left\{ \right. &
    w_1 \mapsto \texttt{w1}, \ldots w_H \mapsto \texttt{wH}, s_1 \mapsto \texttt{s1}, \ldots s_G \mapsto \texttt{sG} \\
      | &\left. \texttt{wi} \in \{1,\ldots,8\} \wedge \texttt{si} \in \{\texttt{unsign},\texttt{sign}\}  \right\}.
\end{align*}
% \gc{I think your typing is off here. It looks like $M$ is the set of maps with a certain domain and range, not one specific one with that type, i.e. whatever the definition is, I guess it should be of the form $M = \{...\}$? Secondly, this function definition looks far too general to me, for example I can change $s_G$, and the bitwidth assigned to $w_1$ can change as a result. Finally I'm unclear about why $w_1$ etc are italicised here... should they be? I suspect these are all notational issues. They will also be helped if you say explicitly what's going on in this section, e.g. via the text above. Here's an alternative... not sure if it captures what you meant:}

\change{We enumerate the entire parameter space, $M$, constructing VeriLang expressions $\llbracket lhs \rrbracket_m$ and $\llbracket rhs \rrbracket_m$ for all $m\in M$, and determine, for each $m\in M$, whether these representations are equivalent.} ROVER converts both $\llbracket lhs \rrbracket_m$ and $\llbracket rhs \rrbracket_m$ to Verilog then deploys a commercial RTL equivalence checker (EC). This enables the re-use of the RTL generation framework (see Section~\ref{sec:code_gen}) and defers Verilog semantic interpretation to the commercial tool. Each mapping corresponds to a single lemma, which the EC either proves (true) or disproves (false). These results are stored in a lookup table $T$ such that 
\begin{equation}
T(m)=
\begin{cases}
\textrm{true}, & \textrm{if } \llbracket lhs \rrbracket_m\cong \llbracket rhs \rrbracket_m \\
\textrm{false}, & \textrm{else.}
\end{cases}
\end{equation}
% \gc{I don't think the above definition makes sense. I thought that $T$ was defined {\em only} on the domain $M$, that it was the extrapolating $\phi$ that was defined on a superset. Yet in this definition, $T$ itself is defined on the larger domain, so the distinction between $T$ and $\phi$ is unclear -- they appear to be the same over $M$ (equation 9) but then not clear beyond that. Let's discuss how to rephrase. I suspect it's simply that the domain of $T$ given here is too general?}

\begin{figure}
    \centering
    \begin{tikzpicture}
  [level distance=12mm,
   % every node/.style={fill=red!60,rectangle, rounded corners,inner sep=1pt},
   % level 1/.style={sibling distance=20mm,nodes={fill=red!45}},
   % level 2/.style={sibling distance=10mm,nodes={fill=scred!50}},
   % level 3/.style={sibling distance=5mm,nodes={fill=red!25}}
   level/.style={sibling distance=24mm/#1}
   ]
\node[text width = 8cm] at (0,0) {$
    (+\; w_3\; w_2\; s_2\; (+\; w_2\; w_1\; s_1\; \textbf{a}\; w_1\; s_1\; \textbf{b})\; w_1\; s_1\; \textbf{c})
    \rightarrow 
    (+\; w_3\; w_1\; s_1\; \textbf{a}\; w_2\; s_2\; (+\; w_2\; w_1\; s_1\; \textbf{b}\; w_1\; s_1\; \textbf{c}))$};

  % \node at (3,-1) {$(a_{w_1}^{s_1}+(b_{w_1}^{s_1}+c_{w_1}^{s_1})_{w_2}^{s_2})_{w_3}^{s_3}$};
    \node at (0,-1)[draw=black,rectangle, rounded corners] {$w_2<w_3$}
     child {node[draw=black,rectangle, rounded corners] {$w_1<w_3$}
        child {node[fill=scdgreen!50,rectangle, rounded corners] {T\tiny{(5)}}}
        child {node[draw=black,rectangle, rounded corners] {$s_1$}
            child{node[draw=black,rectangle, rounded corners] {$s_2$}
                child{node[fill=scdgreen!50,rectangle, rounded corners] {T\tiny{(4)}}}
                child{node[fill=scred!50,rectangle, rounded corners] {F}}
            }
            child{node[fill=scdgreen!50,rectangle, rounded corners] {T\tiny{(3)}}}
            }
        }
     child {node[draw=black,rectangle, rounded corners] {$w_1<w_2$}
       child {node[fill=scred!50,rectangle, rounded corners] {F}}
       child {node[draw=black,rectangle, rounded corners] {$s_1$}
            child{node[draw=black,rectangle, rounded corners] {$s_2$}
                child{node[fill=scdgreen!50,rectangle, rounded corners] {T\tiny{(2)}}}
                child{node[fill=scred!50,rectangle, rounded corners] {F}}
            }
            child{node[fill=scdgreen!50,rectangle, rounded corners] {T\tiny{(1)}}}
            }
     };

     % \node at (0,-7) {$(w_2<w_3 \land w_1<w_2 \land s_1) \lor (w_2<w_3 \land w_1<w_2 \land !s_1 \land !s_2)$};
     {
  \node at (0,-8) [scale=0.88]{
    \begin{tabular}{clclclcll}
        $\phi$ & = & & & &\\ 
        (1) & $(w_2<w_3   $ & $\land$ & $w_1<w_2      $ &$\land $& $s_1)$ && &$\lor$\\
        (2) & $(w_2<w_3   $ & $\land$ & $w_1<w_2      $ &$\land $& $!s_1$ &$\land $& $ !s_2)$ & $\lor$\\
        (3) & $(!(w_2<w_3)$ & $\land$ & $w_1<w_3      $ &$\land $& $s_1)$ &&& $\lor$\\
        (4) & $(!(w_2<w_3)$ & $\land$ & $w_1<w_3      $ &$\land $& $!s_1$ &$\land$&$ !s_2)$ &$\lor$  \\
        (5) & $(!(w_2<w_3)$ & $\land$ & $!(w_1<w_3))$ 
    \end{tabular}
  };
};
\end{tikzpicture}
    \caption{A decision tree classifier, which determines whether the restricted associativity of addition rewrite (shown above the tree) is valid (T) or invalid (F). The right/left branch is taken if the condition is true/false. The $s_i$ nodes evaluate to true when $s_i == \texttt{unsign}$. The decision tree corresponds to the sum of product Boolean expression displayed at the bottom of the tree, where each product corresponds to a particular T leaf.}
    \label{fig:decision_tree}
\end{figure}

The lookup table $T$, represents the data from which ROVER learns a condition. The objective is to determine a condition, $\phi$, that can be extrapolated beyond the domain $M$. To achieve this ROVER fits a decision tree classifier~\cite{Bishop2006PatternLearning} to determine a predicate, $\phi$, such that 
\begin{equation}\label{eqn:classifier_correct}
\forall m\in M,\;\phi(m) = T(m). 
\end{equation}
ROVER uses Python's sklearn library implementation to fit a decision tree classifier. The classifier learns based on Boolean features \eqref{eqn:sign_feature}-\eqref{eqn:shift_feature}.
\begin{align}
    i = 1\ldots m, & \quad s_i==\texttt{unsign} \label{eqn:sign_feature}\\
    i,j,k = 1\ldots n, i\neq j\neq k, & \quad w_i == w_j \\ 
    & \quad w_i < w_j \label{eqn:add_feature}\\
    & \quad w_i \pm 1 < w_j  \\
    & \quad w_i + w_j < w_k \label{eqn:mult_feature}\\
    & \quad w_i + 2^{w_j} < w_k \label{eqn:shift_feature}
\end{align}
These features are relevant for the operators supported in VeriLang. For example, \eqref{eqn:add_feature} indicates whether an addition of $w_i$-bit integers stored in a $w_j$-bit signal will retain a carry-out. Similarly, \eqref{eqn:mult_feature} relates to a multiplication of a $w_i$-bit integer and a $w_j$-bit integer stored in a $w_k$-bit signal. Lastly, \eqref{eqn:shift_feature} relates to a $w_i$-bit integer left-shifted by a $w_j$-bit integer stored in a $w_k$-bit signal.

Starting from depth one, ROVER incrementally increases the maximum decision tree depth during the fitting procedure until the generated classifier satisfies \eqref{eqn:classifier_correct}\change{, corresponding to zero classification error on the training set}. In Figure~\ref{fig:decision_tree}, we take a restricted associativity of addition rewrite as an example, where we force the variables $a,b$ and $c$ to have identical bitwidth and signage parameters. This rewrite contains $H=3$ free bitwidth parameters and $G=2$ free signage parameters. \change{The procedure shown in Figure~\ref{fig:rewrite_cond_flow_diagram} generates $|M| = 8^3\times 2^2= 2048$ equivalence checks. The equivalence check results are used to train a decision tree classifier, which achieves perfect classification accuracy at depth four.} The resulting decision tree is shown in Figure~\ref{fig:decision_tree}, where each T (F) leaf corresponds to valid (invalid) rewrite instances.

The decision tree is converted to a Boolean expression in sum of product form, \change{yielding a $\phi$ that satisfies~\eqref{eqn:classifier_correct}}, where only the leaves that are classified as true are retained. The sum of product expression corresponding to the example decision tree is shown in Figure~\ref{fig:decision_tree}. The minimum depth classifier satisfying \eqref{eqn:classifier_correct} corresponds to a condition with the minimal number of products. Even for a relatively simple rewrite such as the unrestricted associativity of addition, there are $H=5$ free bitwidth parameters and $G=4$ free signage parameters. As a result, the fitting process described above generates a depth 9 decision tree classifier. 
% The condition derived from this decision tree is given in the Appendix. 
% \gc{Can you instead say `are provided in the supplementary material'?}

\change{Via the e-matching process \egg~searches the e-graph for expressions matching the left-hand side of a given rewrite, returning a mapping $m$. ROVER evaluates the synthesized $\texttt{cond}$, $\phi(m)$, to determine whether the rewrite can be applied or not. $\phi$ is guaranteed to be necessary and sufficient if the mapping returned by the e-matching process $m\in M$.} For example, applying the rewrite described in Figure~\ref{fig:decision_tree} to an e-graph corresponding to the Verilog shown in Figure~\ref{fig:verilog_rewriting}, e-matching returns two maps $m_1$ and $m_2$ corresponding to the expressions for \texttt{left1} and \texttt{left2} respectively.  
\begin{tikzpicture}
    \node at (0,0) {$m_1 = \left\{ \begin{array}{ll} 
w_3 & \mapsto 9 \\ 
w_2 & \mapsto 8 \\ 
s_2 & \mapsto \texttt{unsign} \\ 
w_1 & \mapsto 8 \\
s_1 & \mapsto \texttt{unsign} 
 \end{array} \right.$};
     \node at (4.5,0) {$m_2 = \left\{ \begin{array}{ll} 
w_3 & \mapsto 9 \\ 
w_2 & \mapsto 9 \\ 
s_2 & \mapsto \texttt{unsign} \\ 
w_1 & \mapsto 8 \\
s_1 & \mapsto \texttt{unsign} 
\end{array} \right.$};
\end{tikzpicture}
Evaluating the \texttt{cond}, $\phi$, shown in Figure~\ref{fig:decision_tree}
\begin{equation}
    \phi(m_1) = \textrm{false} \hspace{2em} \phi(m_2) = \textrm{true}.
\end{equation}
This agrees with the validity statements made in Figure~\ref{fig:verilog_rewriting}.

\change{Since ROVER supports Verilog with signals exceeding 8-bit integers (the limit of the training data), we extrapolate by assuming that the predicate, $\phi$, learnt on training data is valid for the entire domain of feasible bitwidths, which is an infinite space. Even if this assumption is incorrect, false positives, which we did not observe in practice, are detected by the back-end verification, described in Section~\ref{sect:verification}, preventing ROVER from delivering functionally incorrect RTL.}

%%%%%%%%%%%%%%%%%%%%%%%%%%%%%%%%%%%%%%%%%%%%%%%%%%%%%%%%%%%%%%%%%
% EXTRACTION
%%%%%%%%%%%%%%%%%%%%%%%%%%%%%%%%%%%%%%%%%%%%%%%%%%%%%%%%%%%%%%%%%
\section{Extraction and Back-End}\label{sect:extraction}
ROVER applies rewrites to the e-graph until saturation \change{(defined in Section~\ref{subsec:egraphs})} or a user defined iteration limit is reached. The final e-graph contains a set of valid implementations. The extraction process selects a set of e-classes to implement and within these e-classes chooses the best node to implement that particular e-class. ROVER selects the minimum area design according to a theoretical area metric. 

\subsection{Cost Model} \label{sect:area_cost_model}
The theoretical area metric estimates, per operator, the number of two-input gates required to build that operator, as a function of the input and output parameters. For most logical operators the cost metric is fairly simple, but for the arithmetic operators we fix a particular architecture from amongst the various possibilities. These architecture choices are described in Table~\ref{tab:operator_arch} \change{and are representative of operator architectures implemented by commercial synthesis tools~\cite{Zimmermann2009DatapathDesign}}. When at least one operand is constant we use different constant specific costs, as logic synthesis propagates constants throughout a circuit to reduce the number of gates, e.g. constant multiplication.

Having assigned a cost to each operator, the objective is to minimize the sum of the operator costs. Note that by computing theoretical costs for the merging operators, \texttt{SUM}, \texttt{MUXAR} and \texttt{FMA} downstream synthesis optimizations are encoded directly in the cost model. The theoretical cost metric allows ROVER to efficiently evaluate alternative designs in the e-graph. Commercial ASIC high-level synthesis (HLS) tools use call-outs to logic synthesis engines to evaluate different circuit designs~\cite{Cadence2023StratusHLS}. Such an approach is more computationally intensive thus limiting design space exploration. In Section~\ref{sect:eval}, we evaluate the effectiveness of the theoretical cost metric. 

\subsection{Common Sub-Expression Aware Extraction} \label{sect:ilp}
An accurate circuit area model must correctly account for common sub-expressions. For example a circuit to generate $(a+b)\times(a+b)$ should be costed as \texttt{let} $c=a+b$ \texttt{in} $c\times c$. Such a requirement makes extraction a global problem, since an optimal e-node implementation for a given e-class is no longer local, instead it may depend on implementation choices made in other e-classes. \change{The default greedy extraction method in \egg~fails to account for common sub-expression re-use, therefore yielding sub-optimal solutions.} The common sub-expression problem has been solved by casting extraction as an integer linear programming (ILP) problem \cite{Wang2020SPORES:Algebra}.

Let $\mathcal{N}$ denote the set of all nodes, $\mathcal{C}$ denote the set of all e-classes and $E\subseteq \mathcal{N} \times \mathcal{C}$ be the set of e-graph edges. Additionally, let $\mathcal{N}_c$ be the set of nodes in a particular e-class~$c$. For each node $n\in \mathcal{N}$, we associate some cost, $\text{cost}(n)$, based on the theoretical cost metric and a binary variable $x_n \in \{0,1\}$, indicating whether $n$ is implemented in the final RTL. The objective function of the ILP is described in (\ref{eqn: min_cost}). The program constraints ensure that we extract a valid circuit description. The first constraint (\ref{eqn: choose_child}) ensures that at least one node from all children e-classes of a selected node is implemented. The final constraint ensures that for all output expressions found in the set of e-classes $\mathcal{S}$, we generate a circuit producing that output.

\begin{align}
    \text{minimize: } \sum_{n \in \mathcal{N}} \text{cost}(n)x_n \text{ subject to:} \label{eqn: min_cost}\\
    \forall (n,c) \in E. \; x_n \leq \sum_{n' \in \mathcal{N}_c} x_{n'} \label{eqn: choose_child}\\
    \forall c \in {\mathcal S}. \; \sum_{n \in \mathcal{N}_c} x_n = 1.\label{eqn: all_outputs}
\end{align}
Since e-graphs may contain cycles we include additional topological sorting variables associated with each class $t_c$. Let $N$ denote the number of e-classes and $\mathcal{C}(n)$ be the e-class containing node $n$. The constraint (\ref{eqn:topo_sort}) ensures that the output expression is acyclic.
\begin{equation} \label{eqn:topo_sort}
    \forall (n,k) \in E\quad t_{\mathcal{C}(n)} - Nx_n - t_k \geq 1 - N
\end{equation}

Selecting a node $n\in \mathcal{N}_c$ with child $k$, {\em i.e.}~$x_n = 1$, the constraint simplifies to $t_c\geq t_k + 1$ to get a topologically sorted result, whereas in the case $x_n = 0$, the constraint is vacuously satisfied. To solve this ILP problem we deploy the CBC solver~\cite{Forrest2023Coin-or/Cbc:Releases/2.10.10}. The ILP solution corresponds to a single VeriLang expression, that is a minimal circuit implementation according to the theoretical area metric. 

\subsection{Code Generation}\label{sec:code_gen}
Having obtained a VeriLang expression, ROVER translates this expression into System Verilog to be processed by downstream synthesis tools. The translation is implemented as an e-class analysis, as described in Section \ref{subsec:egraphs}. Initializing a code generation e-graph with a single VeriLang expression, the e-class analysis is constructed from the leaves upwards producing a valid System Verilog implementation. To each e-class we assign a unique signal name, its defined bitwidth and the System Verilog string that implements the particular operation in the e-class. Each e-class in the e-graph corresponds to a single line of functional System Verilog in the output. Traversing the e-graph, ROVER defines a signal at each e-class and assigns the stored expression to that signal name. 

An advantage of the e-graph approach is that ROVER can maintain a mapping between user defined signal names and e-classes throughout the exploration. If such an e-class is present in the extracted implementation, ROVER overwrites the signal name of the appropriate e-class in the code generation e-graph. As a result, the generated System Verilog retains a subset of the original signal names. For example, if a user defined a signal \texttt{two\_x}, assigning it to the expression $x+x$, and that was rewritten as $x\ll 1$, then the \texttt{two\_x} signal would still appear in the generated output, with a different assignment.

%%%%%%%%%%%%%%%%%%%%%%%%%%%%%%%%%%%%%%%%%%%%%%%%%%%%%%%%%%%%%%%%%
% VERIFICATION
%%%%%%%%%%%%%%%%%%%%%%%%%%%%%%%%%%%%%%%%%%%%%%%%%%%%%%%%%%%%%%%%%
\section{Verification}\label{sect:verification}
To increase trust and ensure that the input and generated circuit designs are equivalent, ROVER generates verification scripts for a commercial EC. In many cases, the EC is able to prove the functional equivalence of the input and ROVER generated RTL, without any additional guidance. However, there are instances where the equivalence engine returns an inconclusive result~\cite{Koelbl2009SolverChecking}. Debugging inconclusive proofs can be time consuming for verification engineers. To provide a robust verification flow, ROVER uses the \egg~proof production feature~\cite{Flatt2022SmallClosure} described in Section~\ref{subsec:egraphs}, to decompose the verification problem into a sequence of simple sub-problems.

ROVER uses proof production to extract a sequence of intermediate VeriLang expressions, differing by a single local rewrite at each step. The sequence traces a path between the input and optimized expressions, as shown in Figure~\ref{fig:flow_diagram}. Using the ROVER back-end, each intermediate VeriLang expression is converted to System Verilog. Each pair in the sequence is proven equivalent using the EC, constructing the chain of reasoning that the original and optimized implementations are equivalent. To further aide proof convergence, ROVER identifies the specific signal modified in each pair via an additional lemma. ROVER's proof sequences can contain hundreds of intermediate steps. ROVER generates both the RTL and proof scripts, providing a proof certificate to the user which can be re-run to verify the RTL.

% \begin{figure}
  % \input{mcm_proof_len.tex}
  % \caption{Dataflow graph of an optimized multiple constant multiplication circuit design generated by {\egg}.}
  % \label{fig:mcm_proof_length}
% \end{figure}

%%%%%%%%%%%%%%%%%%%%%%%%%%%%%%%%%%%%%%%%%%%%%%%%%%%%%%%%%%%%%%%%%
% RESULTS
%%%%%%%%%%%%%%%%%%%%%%%%%%%%%%%%%%%%%%%%%%%%%%%%%%%%%%%%%%%%%%%%%
\section{Results}\label{sect:results}
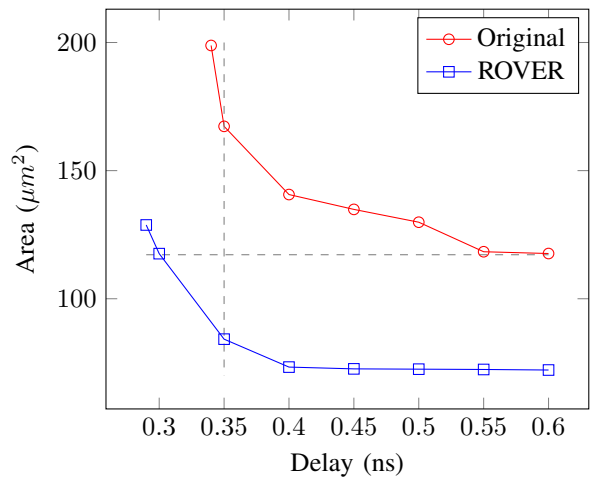
\begin{figure}
    \centering
    \begin{tikzpicture}[]
	\begin{axis}[%
		xlabel=Delay (ns),ylabel=Area ($\mu m^2$)]

\addplot[
    color=red,
    mark=o,
    ]
    coordinates {
(0.34,198.841860)
(0.35,167.279490)
(0.4,140.665141)
(0.45,134.871031)
(0.5,129.869461)
(0.55,118.345500)
(0.6,117.649350)
% (0.94,105)
};
\addlegendentry{Original}

\addplot[
    color=blue,
    mark=square,
    ]
    coordinates {
(0.29,128.787750)
(0.30,117.6)
(0.35, 84.234150)
(0.4, 73.331370)
(0.45, 72.635220)
(0.5, 72.517410)
(0.55, 72.399600)
(0.6, 72.185400)
% (0.65, 68.276251)
% (0.7, 67.772881)
    };
    \addlegendentry{ROVER}

    \addplot[mark=none, gray, dashed] coordinates {
        (0.29, 117.178)(0.6, 117.178)};

    \addplot[mark=none, gray, dashed] coordinates {
        (0.35, 200)(0.35, 70)};

    \end{axis}
\end{tikzpicture}
    \caption{Area-delay profiles for the original and ROVER optimized Media Kernel designs. The dashed grey lines indicate the minimum area and delay comparison points used in Table~\ref{tab:results_table}. }
    \label{fig:media_kernel_profile}
\end{figure}

We used ROVER to optimize a number of industrially and academically sourced RTL benchmarks, automatically producing optimized RTL implementations. The original and optimized designs are synthesized using a commercial synthesis tool for a TSMC 5nm cell library. We also study the synthesis reports to analyze the effectiveness of ROVER's datapath clustering optimizations. Using the approach described in Section \ref{sect:verification} we verified the functional equivalence of the original and optimized architectures. 
\change{We compare each pair of designs at two points along the area-delay trade-off curve using logic synthesis. Firstly, we compare at the minimal delay target at which both designs can meet timing (rounded to the nearest 10 picoseconds), corresponding to the vertical dashed line in Figure~\ref{fig:media_kernel_profile}. The second comparison point, is at the minimum area that both designs can fit within (yielding different performance levels), corresponding to the horizontal dashed line in Figure~\ref{fig:media_kernel_profile}.}

The results are summarized in Table \ref{tab:results_table}. \change{We will primarily focus on the area and delay impact since the cell count and power measurements are proportional to the area in this work.} In Figure~\ref{fig:media_kernel_profile} we plot the area-delay profile comparing the original and ROVER optimized designs across the delay spectrum. We separate the results into two contributions. Firstly, we show how ROVER can optimize general RTL benchmarks. Then we demonstrate how ROVER can optimize different instances of parameterizable RTL, generating a suite of tailored implementations.

\begin{table*}[t]
    \centering
    \caption{Logic synthesis results comparing the original and ROVER optimized designs under two different synthesis constraints. Firstly, at the minimum delay which both designs could meet and secondly, constrained to the minimum area that both designs could meet. Delay, power and area are measured in ns, $\mu W$ and $\mu m^2$, respectively. We bold the best result for each metric.}
    \scalebox{0.95}{
    \begin{tabular}
    {l l
     c r r r r r r
     r r r
     }
        \toprule
        \multirow{2}{*}{Source} & \multirow{2}{*}{Benchmarks} & \multirow{2}{*}{Min Delay}  & \multicolumn{3}{c}{Original} & \multicolumn{3}{c}{ROVER} & \multirow{2}{*}{Min Area} & \multicolumn{1}{c}{Original}  & \multicolumn{1}{c}{ROVER}\\
         \cmidrule(lr){4-6}
         \cmidrule(lr){7-9}
         \cmidrule(lr){11-11}
         \cmidrule(lr){12-12}
        & &  & \multicolumn{1}{c}{Cells} & \multicolumn{1}{c}{Power} & \multicolumn{1}{c}{Area} & \multicolumn{1}{c}{Cells} & \multicolumn{1}{c}{Power} & \multicolumn{1}{c}{Area} &  & \multicolumn{1}{c}{Delay} & \multicolumn{1}{c}{Delay}\\
                 \cmidrule(lr){1-2}
                 \cmidrule(lr){3-9}
                 \cmidrule(lr){10-12}
        % \hline
        \multirow{2}{*}{Intel} & Media Kernel           & 0.35  & 1759 & 959.4 & 167.3 & \bst{918}  & \bst{427.9} & \bst{84.2}  (\texttt{-}50\%) & 117.6  & 0.60 & \bst{0.30} (\texttt{-}50\%)\\
                               & Weight Calculation     & 0.25  & 1353 & 927.1 & 75.3  & \bst{1030} & \bst{719.4} & \bst{ 57.8} (\texttt{-}23\%) & 39.8   & 0.84 & \bst{0.40} (\texttt{-}52\%)\\
                \cmidrule(lr){1-2}
                 \cmidrule(lr){3-9}
                 \cmidrule(lr){10-12}
        \multirow{5}{*}{Open-Source}& FIR Filter Kernel  & 0.67 & 8067 & 2839.0 & 552.6 & \bst{7846}     & \bst{1837.9}  & \bst{428.6} (\texttt{-}22\%) & 209.0  & 4.40 & \bst{4.09} (\texttt{-}07\%) \\
        & ADPCM Decoder \cite{Lee1997MediaBench:Systems} & 0.12 & 620  & 197.4  & 41.8  & \bst{556}      & \bst{190.6}   & \bst{38.0}  (\texttt{-}09\%) & 20.8   & \bst{0.84} & \bst{0.84} (\texttt{+}00\%) \\ 
        & Shifted FMA                                    & 0.22 & 996  & 502.0  & 83.7  & \bst{855}      & \bst{445.1}   & \bst{68.6}  (\texttt{-}18\%) & 54.6   & 0.85 & \bst{0.31} (\texttt{-}64\%) \\
        & Shift Mult                                     & 0.30 & 2864 & 1356.4 & 240.1 & \bst{1317}     & \bst{522.0}   & \bst{88.8}  (\texttt{-}63\%) & 150.7  & 1.88 & \bst{0.26} (\texttt{-}86\%) \\
        & MCM(3,7,21)                                    & 0.12 & \bst{894}  & \bst{161.0}  & \bst{36.6} & 1015 & 249.2  & 51.4  (\texttt{+}40\%) & 23.3   & 0.81 & \bst{0.58} (\texttt{-}28\%)  \\
        & MCM(5,93)                                      & 0.12 & \bst{687}  & \bst{204.8}  & \bst{38.2} & 778  & 292.0  & 53.6  (\texttt{+}40\%) & 22.4   & 0.73 & \bst{0.58} (\texttt{-}21\%)  \\
        & MCM(7,19,31)                                   & 0.09 & \bst{1079} & \bst{230.0}  & \bst{53.3} & 1082 & 236.4  & 54.1  (\texttt{+}02\%) & 21.8   & \bst{0.72} & \bst{0.72} (\texttt{-}00\%)  \\
        \bottomrule
    \end{tabular}
}

    \label{tab:results_table}
\end{table*}

\begin{table}
    \centering
    \caption{ROVER performance and e-graph size before/after rewriting.}
    \begin{tabular}{l r r c r}
        \toprule
        Benchmark         & Init Nodes & Final Nodes & Extract & Runtime (sec) \\
        \midrule
        Media Kernel             & 45  & 1312  & ILP & 10.67 \\
        Weight Calc.             & 107 & 3036  & ILP & 165.00 \\
        FIR Filter               & 30  & 8487  & ILP & 155.90 \\ 
        ADPCM                    & 17  & 7290  & Greedy & 16.64 \\
        Shifted FMA              & 13  & 26    & ILP & 0.09 \\
        Shift Mult               & 13  & 72    & ILP & 0.13 \\
        MCM(3,7,21)              & 13  & 17493 & ILP & 135.00 \\
        MCM(5,93)                & 12  & 2986  & ILP & 113.86 \\
        MCM(7,19,31)             & 13  & 7601  & ILP & 50.59 \\
        \bottomrule
    \end{tabular}
    \label{tab:benchmark+ops}
\end{table}

\subsection{Exploiting Datapath Optimizations}
The first set of benchmarks in Table~\ref{tab:results_table} are Intel RTL designs. The first benchmark is a kernel from the Intel media module. The initial design was optimized by hand by \change{a hardware design expert}. ROVER is able to automatically optimize the design and achieve comparable results to the manually optimized RTL, discovering the opportunity to merge two multiplication arrays into a single array using the ``Merge Mult Array'' rewrite. Studying the reports generated by the synthesis tool, we can identify the source of the area reduction. The original design produces four datapath clusters, corresponding to four carry-propagate adders in the synthesized netlist. By contrast, the ROVER optimized design produces two datapath clusters, halving the number of carry-propagate adders in the generated netlist. These improvements translate to a 14.7\% reduction in minimum achievable delay within a circuit area 35.4\% smaller. \change{In the logic synthesis engine, further arithmetic clustering is prevented because the tool detects datapath leakage (as described in Section~\ref{subsect:datapath_opto}) due to supposed truncation in the following System Verilog.
\begin{equation*}
\texttt{a[8:0]= 9'd256 - \{1'b0,b[7:0]\};}
\end{equation*}
This analysis, however, is flawed. There is in fact no overflow as we are dealing with constants. ROVER meanwhile, rewrites this expression to avoid this supposed datapath leakage.}
The Weight Calculation benchmark is a two-stage pipelined design computing pixel offsets in the graphics pipeline. ROVER optimizes each stage independently. By rewriting the MUX tree structure within each stage, using the ``Sel Mul'' rewrites, ROVER reduces the number of multipliers instantiated from five to three. \change{The work of Verma {\em et al.}~\cite{dataflow2008verma} has no ability to combine multipliers by manipulating the MUX tree structure, so can not reach these designs generated by ROVER.}

The next two benchmarks are taken from~\cite{dataflow2008verma}, where ROVER generalizes and exceeds the capabilities of this prior work. The first example is a familiar finite impulse response (FIR) filter with 8-taps (a 3-tap version is shown in Figure~\ref{fig:fir_4_arch_0}). Via the ``Arithmetic Logic Exchange'' rewrites, ROVER explores all the alternative arithmetic clustering opportunities extracting an optimal clustering according to the theoretical cost metric. \change{In contrast, the logic synthesis engine appears to greedily cluster all operators. This maintains carry-save representation throughout, but, we speculate, results in shifting carry-save representations, incurring additional circuit area overhead.} The ADPCM decoder is a design which approximates a $16\times 4$ multiplier. \change{For this benchmark, both ROVER and the logic synthesis engine achieve a complete clustering. ROVER manipulates the MUX tree structure, whilst the logic synthesis tool appears to add additional operators to facilitate the clustering.}

The next two benchmarks demonstrate optimizations beyond the capabilities of \cite{dataflow2008verma}. Shifted FMA exploits multiplication-manipulating rewrites since logic synthesis tools will effectively cluster multiplications followed by additions to reduce the number of carry-propagate adders. \change{As in the FIR filter example, the logic synthesis greedily clusters, such that it must perform a shift of a carry-save representation. By moving the shift ROVER enables a simpler arithmetic clustering.} Shift Mult is a kernel extracted from a floating point multiplier that normalizes the product of two denormals. By re-ordering the shift and multiplication operators a smaller multiplier can be instantiated, reducing the circuit area. \change{In contrast, the logic synthesis tool does not manipulate the higher-level dataflow graph to explore the interaction of arithmetic and logical operators, and does not discovers this opportunity. These ROVER optimizations are not reachable by~\cite{dataflow2008verma}, since their tool did not explore the interaction between multiplication and logic. }

The ``Constant Expansion'' rewrites are valuable for the MCM benchmarks, \change{where for MCM($a_1$,$a_2$,...,$a_n$) we ask ROVER to generate optimized RTL producing $\{a_1\times x, a_2\times x,..., a_n \times x\}$. MCM(3,7,21) is an example taken from \cite{Gustafsson2007AProblems}.} ROVER is able to match the operator count from \cite{Gustafsson2007AProblems}, extracting a design that uses three addition/subtraction operators by sharing intermediate results. Such an architecture serializes the construction of $3\times x$ and $21 \times x$, which at low delay targets introduces an area penalty, because the original architecture can compute each result in parallel with no dependency. However, from the ROVER generated RTL a smaller circuit can be synthesized, as shown in Table~\ref{tab:results_table}. \change{For the MCM(5,93) benchmark ROVER is similarly able to use just 3 adders, matching the minimal adder count, and showing similar synthesis results to MCM(3,7,21). For the MCM(7,19,31) benchmark\footnote{Thank you to the anonymous reviewer for providing this benchmark.} ROVER recovers the standard CSD solution using 4 adders and matching the synthesis tool (hence the identical synthesis results). The minimal solution uses 3 adders, but is unreachable using ROVER's existing rewrites as it relies upon representing $19=(31+7)\gg 1$.}

\change{In this work, we used the logic synthesis tool with all datapath optimizations enabled to provide a baseline. However, this baseline includes state-of-the-art datapath optimization techniques. If we disable these optimizations we get an alternative baseline that highlights the significance of the datapath optimizations built-in to the logic synthesis and those performed by ROVER. On average, with datapath optimization disabled the logic synthesis tool produced circuits \avgareavsls~larger than with datapath optimization enabled, and \avgareavsrov~larger than the ROVER generated circuits. Furthermore, in 5 out of the 9 benchmarks, disabling datapath optimization led to timing violations in the synthesized netlists. }

\subsection{Bitwidth Dependent Architectures}
In this section we consider parameterizable RTL designs. As the complexity of integrated circuits grows, reusable and parameterizable hardware has become increasingly popular amongst engineers and architects as it facilitates faster development. Each instance of this RTL will be synthesized using the same architecture. By contrast, ROVER automatically optimizes each instance generating a bespoke component that is optimized for a given instance. 

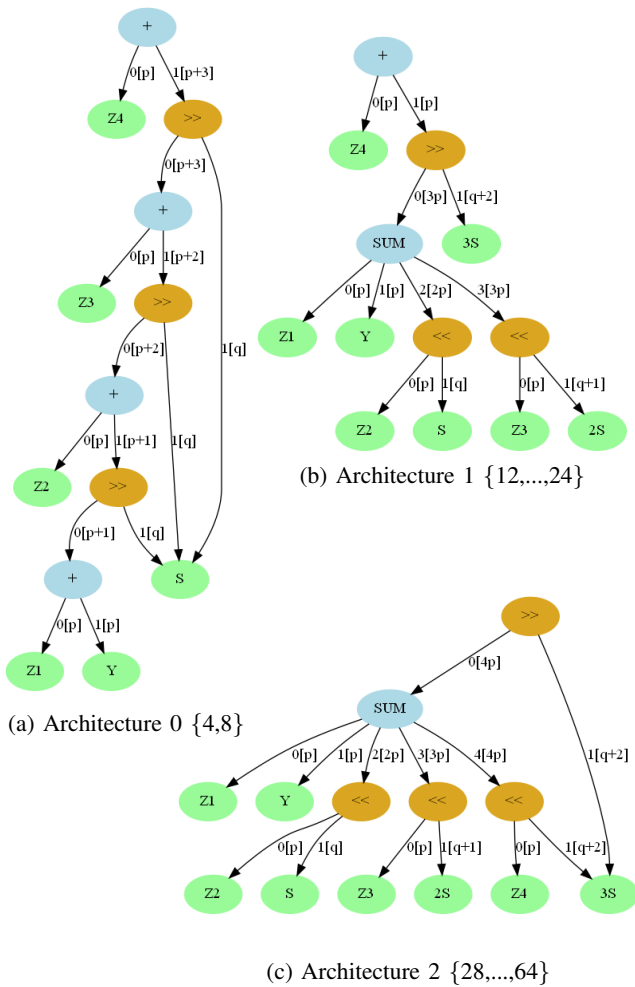
\begin{figure}
     \centering    
     \begin{tikzpicture}
        \node at (4.2, 1.4) {\resizebox{5cm}{!}{\includegraphics{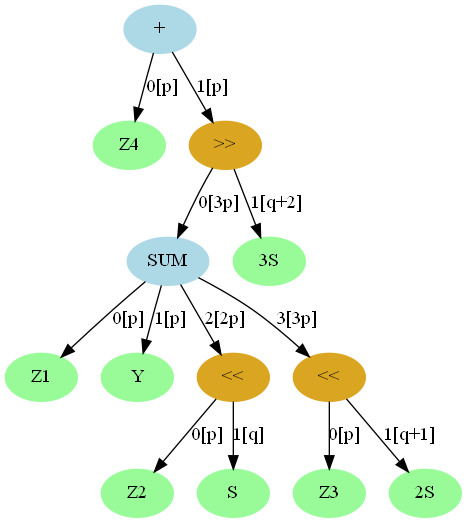}\label{fig:fir_4_arch_1}}};
        \node at (3.75, -5.4) {\resizebox{6.2cm}{!}{\includegraphics{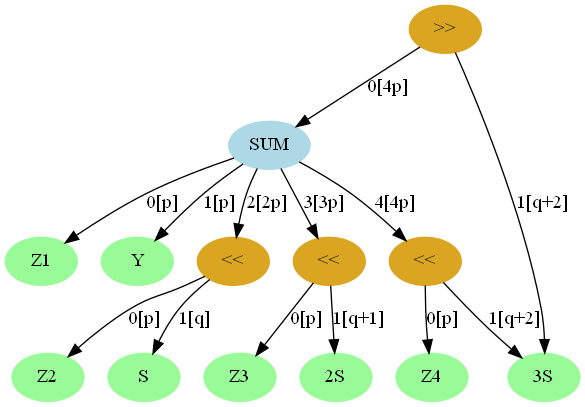}\label{fig:fir_4_arch_2}}};
        \node at (0,0) {\resizebox{3.3cm}{!}{\includegraphics{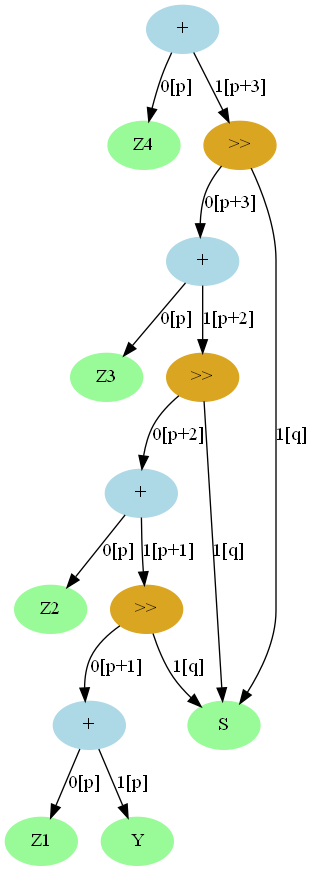}\label{fig:fir_4_arch_0}}};
        \node at (0, -5) {\small (a) Architecture 0 \{4,8\}};
        \node at (4.2, -1.7) {\small (b) Architecture 1 \{12,...,24\}};
        \node at (3.75, -8.3) {\small (c) Architecture 2 \{28,...,64\}};
  \end{tikzpicture}
     \caption{Simplified FIR filter data-flow graphs representing optimal architectures for different choices of the input bitwidth parameter $p$ and shift bitwidth parameter $q$. Edge labels indicate the operator index and bitwidth in square brackets. The sets in curly braces are bitwidths for which that architecture is optimal. 
     In these graphs $2S$ and $3S$ are constant multiples of $S$.
     }
     \label{fig: 4 tap fir}
 \end{figure}

To investigate whether ROVER can usefully adapt the architecture depending on parameter values, we considered a 3-tap FIR filter with parameterizable input bitwidths. We passed ROVER each design, increasing the input bitwidth parameter from 4 to 64 and allowed ROVER to explore the design space for each parameterization. As shown in Figure~\ref{fig: 4 tap fir}, ROVER extracted one of three distinct architectures. In the FIR kernel testcase the benefits of clustering consecutive additions into a \texttt{SUM} node compete with the additional shift operations required to facilitate the merging. Note that Architecture 0 uses four carry-propagate adders, Architecture 1 uses two carry-propagate adders, whilst Architecture 2 uses only a single carry-propagate adder at the expense of additional shifting logic. ROVER automatically detects the point at which this tradeoff becomes favourable. 

\begin{figure}
    \centering
    \begin{tikzpicture}
\begin{axis}[
	% enlargelimits=0.05,
	% legend style={at={(0.5,0.15)},
	% anchor=north,legend columns=-1},
	% ybar interval=1,
    % ybar,
    % xtick={8,16,24,32,40,48,56,64},
    scatter,
    only marks,
    xlabel={Bitwidth},
    ylabel={Area Change (\%)},
    ymajorgrids=true,
    xmajorgrids=true,
    % symbolic x  coords={4,8,12,16,20,24,28,32,36,40,44,48,52,56,60,64},
    % symbolic x coords={
    %        4,8,12,16,20,24,28,32,36,40,44,48,52,56,60,64},
    xtick={
            8,16,24,32,40,48,56,64},
    % nodes near coords
    grid style=dashed,
]

\addplot[
        scatter,only marks,scatter src=explicit symbolic,
        scatter/classes={
            a={mark=square*,blue},
            b={mark=triangle*,red},
            c={mark=o,draw=black,fill=black}
        }
    ]
    table[x=x,y=y,meta=label]{
        x    y    label
        4  10.5628373 a
        8  13.6329305 a
       12  1.3741769  b
       16 -22.6809053 b
       20 -17.9390512 b
       24 -25.2490207 b
       28 -11.5701571 c
       32 -20.4329372 c
       36 -08.4886128 c
       40 -17.8021499 c
       44 -23.8176188 c
       48 -22.2253665 c
       52 -20.7342114 c
       56 -26.3397896 c
       60 -22.8977497 c
       64 -30.1162138 c
    };

\legend{Arch 0, Arch 1, Arch 2}
\end{axis}
\end{tikzpicture}
    \caption{Synthesis results for the 3-tap FIR kernel at a range of different bitwidths. We synthesized both the ROVER generated RTL and original RTL (Architecture 0) with a minimum delay objective. We plot the relative change in area and delay against the baseline.}
    \label{fig:fir_3_data}
\end{figure}
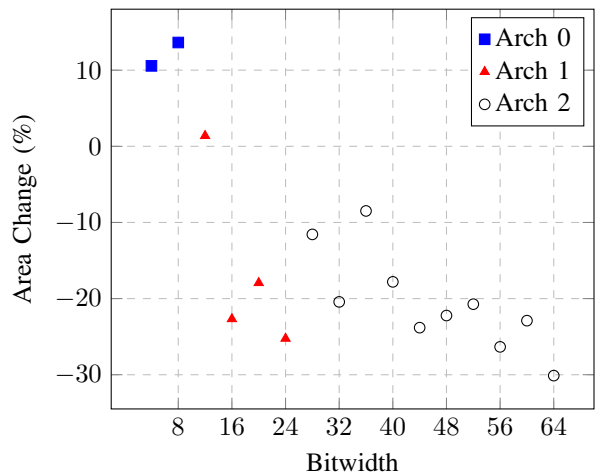

For each bitwidth, we synthesized Architecture 0 and the distinct ROVER generated RTL (which implements either Architecture 0, 1 or 2) at the minimum delay target that both can meet. Figure \ref{fig:fir_3_data} plots synthesis results at each bitwidth comparing against the baseline, Architecture 0 (Figure \ref{fig: 4 tap fir}). The architectural selections made by ROVER reduce the circuit area by up to 30\% and by 15\% on average. For 4-bit and 8-bit designs, ROVER increases the circuit area despite deploying the same architecture as the baseline. This is due to synthesis noise, an effect quantified precisely in~\cite{Coward2022AutomaticE-Graphs}. Using ROVER to automatically generate an optimized design for each parameterization allows engineers to avoid manual customization without sacrificing IP quality.

\subsection{Performance}
Table~\ref{tab:benchmark+ops} presents benchmark properties and optimization statistics. For the ILP extraction method, we set a timeout limit of 120 seconds and in all the longer running benchmarks, ILP solving dominated the runtime. \change{Note that the number of ILP constraints is proportional to the number of nodes in the final e-graph. Whilst ILP scalability is a concern, the modular nature of RTL design ensures that we rarely meet large scale problems.} We resorted to the faster greedy \egg~extraction method~\cite{Willsey2021Egg:Saturation} for the ADPCM decoder since there was no scope to exploit common sub-expressions in this benchmark. Extraction method selection is a user defined option for ROVER. We note that the final e-graph size is not well correlated with the number of operators in the initial e-graph. The size of the final e-graph depends more upon the structure of the initial design.

Highlighting the importance of the verification flow, for the Media Kernel and Shift Mult benchmarks, the EC returned inconclusive results, even when running for several hours, when only given the original and ROVER generated RTLs. Using the ROVER generated problem decomposition, the correctness of the generated RTL could be proven in seconds. For all other benchmarks presented here, the EC could prove the equivalence of the original and ROVER generated RTLs without the problem decomposition described in Section~\ref{sect:verification}. 

%%%%%%%%%%%%%%%%%%%%%%%%%%%%%%%%%%%%%%%%%%%%%%%%%%%%%%%%%%%%%%%%%
% COST METRIC EVALUATION
%%%%%%%%%%%%%%%%%%%%%%%%%%%%%%%%%%%%%%%%%%%%%%%%%%%%%%%%%%%%%%%%%
\section{Cost Metric Evaluation}\label{sect:eval}
The primary objective of the theoretical cost metric is to steer the extraction process in order to generate an optimized architecture. Previously, we evaluated the noise floor in logic synthesis to understand inherent variability of such a complex tool~\cite{Coward2022AutomaticE-Graphs}. \change{We used an approach known as performance fuzzing~\cite{Chen2020ATesting,Zhou2023Mariposa:Verification}, that differs from the more traditional application of fuzzing to automated bug detection~\cite{Chen2020ATesting}. We randomly applied non-functional mutations to designs, for example renaming a variable in RTL, and observed up to a 15\% difference in logic synthesis area.} ROVER's cost model cannot be expected to capture \change{this. The variability is equally likely to benefit ROVER as it is to be detrimental for the results shown in Table~\ref{tab:results_table}. However, the overall benefit demonstrated by ROVER is statistically significant and explainable.}

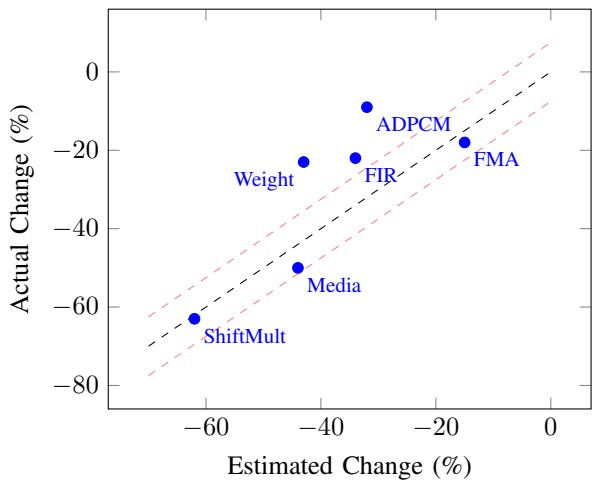
\begin{figure}
    \centering
    \begin{tikzpicture}
 
\begin{axis}[xlabel=Estimated Change (\%),ylabel= Actual Change (\%)]
    \addplot[
    color=black,
    dashed
    ]
    coordinates {
    (-70,-70)(0,0)
    };

    \addplot[
    color=red!50,
    dashed
    ]
    coordinates {
    (-70,-77.5)(0,-7.5)
    };

    \addplot[
    color=red!50,
    dashed
    ]
    coordinates {
    (-70,-62.5)(0,7.5)
    };
    
    \addplot[%
    scatter/classes={a={blue}, b={red}},
        scatter, mark=*, only marks, 
        scatter src=explicit symbolic,
        nodes near coords*={\Label},
         every node near coord/.append style={anchor=north west, font=\footnotesize},
         % nodes near coords align={below,right},
        visualization depends on={value \thisrow{label} \as \Label},
        color=blue
    ] table [meta=class] {
x y class label
-34 -22 a FIR
-44 -50 a Media
-32 -09 a ADPCM
-15 -18 a FMA
-62 -63 a ShiftMult
% -43 -23 a Weight
    };

\addplot[mark=*, color=blue, nodes near coords={\footnotesize Weight}, every node near coord/.style = {anchor=north east, align=center, color=blue}, only marks]
 coordinates {(-43, -23)};
    
\end{axis}
\end{tikzpicture}
    \caption{ROVER's predicted percentage change vs. the actual percentage change based on logic synthesis at the minimum delay target. Points above/below the diagonal indicate that ROVER over/under-predicts the area reduction. We omit the MCM results. Red lines represent the synthesis noise window.}
    \label{fig:cost_metric_comparison}
\end{figure}

To evaluate the accuracy of the cost model, we plot the ROVER estimated circuit area reduction against the actual change seen in the logic synthesis results at the minimum delay target in Figure~\ref{fig:cost_metric_comparison}. The graph shows that ROVER both under- and over-estimates the benefit of its optimizations but does provide a reasonable indicator. The ADPCM and Weight benchmarks exhibit significant over-estimates. In the ADPCM example, ROVER manipulates the MUX tree structure of the design to enable arithmetic clustering, which the synthesis tool exploits successfully. Analyzing the datapath extraction report generated during synthesis of the original ADPCM design, we see that the synthesis tool is already capable of manipulating this design to cluster the arithmetic operations limiting the observable benefit of ROVER's optimizations. For the Weight Calculation benchmark, ROVER reduces the number of multipliers instantiated by two. In the original design, the synthesis tool includes these multipliers in a datapath cluster, therefore the circuit area benefit is less than the full multiplier area cost. The omitted MCM benchmarks highlight the limitations of an area only model, as the benefit depends upon the delay target. 

\section{Conclusion}
This paper presents methods to exploit the properties of the e-graph data structure, finding an ideal application in the RTL optimization problem.
E-graphs greatly simplify this problem by avoiding any need to specify an order in which to apply transformations whilst maintaining bit identical functionality. The e-graph's foundations rest on functional equivalence principles, which are crucial in 
hardware design where the correctness requirements are higher than most other domains. By defining a set of parameterized bitvector-manipulating
transformations, learnt from Intel engineers, we have matched human-engineered designs in terms of circuit quality. The productivity and circuit quality benefits that stem from automated rewriting techniques allow engineers to write behavioural, less bug prone designs and leave the optimization to a tool that can provide verified implementations. 

Future work will seek to address delay optimization; this will allow ROVER to select different arithmetic operator architectures depending on the timing budget available. 
\change{We will also address the limitations of the rewrite condition synthesis flow, which currently relies upon an unproven extrapolation assumption. We believe the integration of a theorem prover such as ACL2~\cite{Hunt2017IndustrialACL2} will allow us to prove this assumption. For extraction, we will resolve the ILP bottleneck in ROVER's current implementation, by leveraging the outcome of a community effort to improve common sub-expression aware extraction\footnote{https://github.com/egraphs-good/extraction-gym}. Lastly, as noted in the MCM discussion, there are scenarios in which bespoke tools yield optimal solutions more efficiently. Through dynamic rewrites, we will provide an interface to such tools.}

\ifCLASSOPTIONcompsoc
  \section*{Acknowledgments}
\else
  \section*{Acknowledgment}
\fi

The authors would like to thank Yann Herklotz for the fuzzing tool and Bryan Tan who contributed useful rewrites.

\ifCLASSOPTIONcaptionsoff
  \newpage
\fi



\bibliographystyle{IEEEtran}
\bibliography{IEEEabrv,references}
%


\begin{IEEEbiography}[{\includegraphics[width=1in,height=1.25in,clip,keepaspectratio]{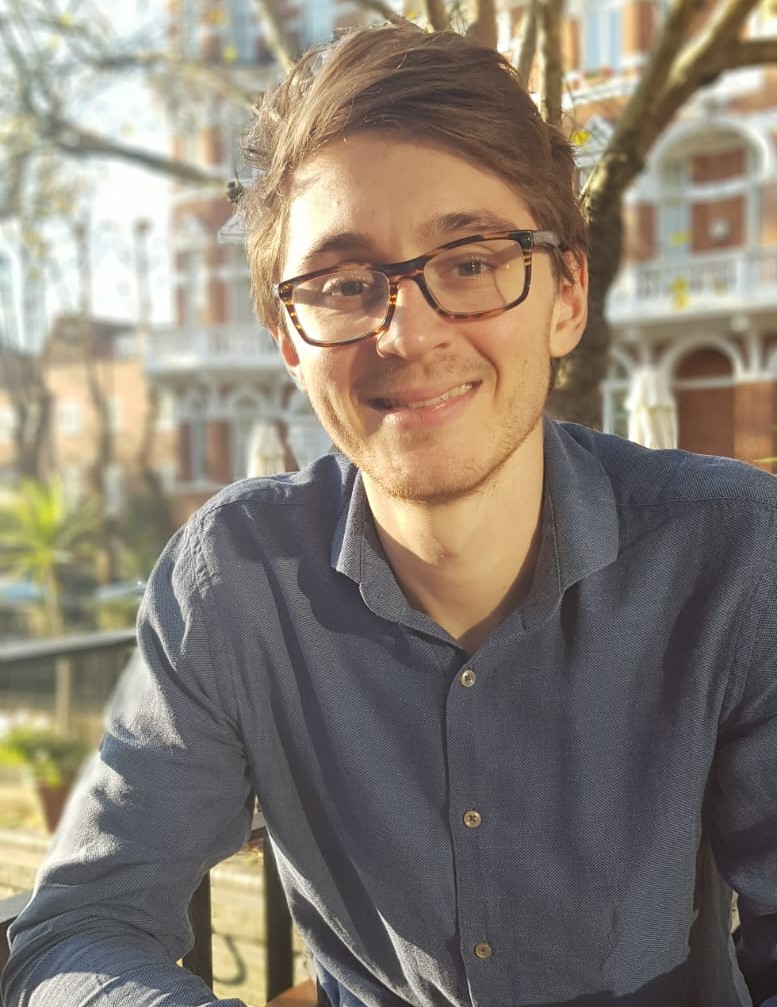}}]{Samuel Coward}
received a BSc in Mathematics in 2018 and an MPhil in Scientific Computing in 2019, from the University of Cambridge. He is currently studying for a PhD in Electrical and Electronic Engineering at Imperial College London, whilst also working as a Graphics Hardware Engineer at Intel Corporation. Samuel's research focuses on automating RTL design and program analysis techniques to increase chip design productivity and quality. 
\end{IEEEbiography}

\begin{IEEEbiography}
[{\includegraphics[width=1in,height=1.25in,clip,keepaspectratio]{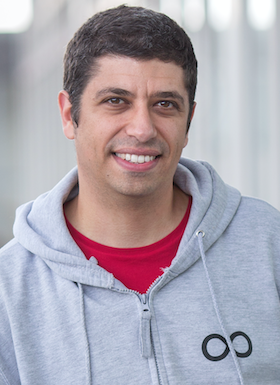}}]
{George A. Constantinides} (S'96, M'01, SM'08) received the Ph.D. degree from Imperial College London in 2001. Since 2002, he has been with the faculty at Imperial College London, where he is currently Professor of Digital Computation and Associate Dean of Engineering. He has served as chair of the FPGA, FPL and FPT conferences. He currently serves on several program committees and has published over 200 research papers in peer refereed journals and international conferences. Prof.~Constantinides is a Senior Member of the IEEE and a Fellow of the British Computer Society. 
\end{IEEEbiography}

%\newpage

\begin{IEEEbiography}
[{\includegraphics[width=1in,height=1.25in,clip,keepaspectratio]{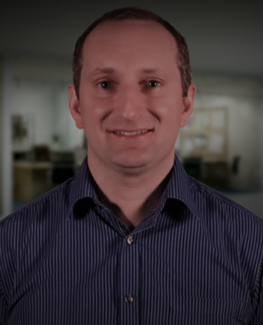}}]{Theo Drane}
started working for the Datapath consultancy, Arithmatica, in 2002 after a Mathematics degree from the University of Cambridge, UK. He moved to Imagination Technologies in 2005, where he subsequently founded their Datapath team while studying for a PhD at Imperial College London's EEE Department. In December 2018, after a stint within Cadence Design System’s Logic Synthesis division, Genus, he joined Intel’s Graphics Group. His applied research Numerical Hardware \& System Level Design Group focuses on all aspects of architecting, implementing, optimizing and verifying math intensive hardware.
\end{IEEEbiography}


%\vfill


\end{document}